\documentclass[%
 reprint,
 amsmath,amssymb,
 aps,
 pre,
]{revtex4-1}

\usepackage{graphicx}
 \graphicspath{{./}
{./figures/}
}
\usepackage{dcolumn}
\usepackage{bm}
\usepackage{color}
\usepackage{hyperref}
\newcommand{\mage}[1]{\textcolor{black}{#1}}
\newcommand{\blue}[1]{\textcolor{black}{#1}}
\newcommand{\red}[1]{\textcolor{black}{#1}}

\begin{document}

\preprint{APS/123-QED}


\title{Numerical solutions of Fokker-Planck equations with drift-admitting jumps}


\author{Yaming Chen}
\email[]{chenym-08@163.com}
\author{Xiaogang Deng}
\affiliation{College of Aerospace Science and Engineering, National University of Defense Technology, Changsha 410073, China}


\date{\today}

\begin{abstract}
      We develop a finite difference scheme based on a grid staggered by flux points and solution points to solve Fokker-Planck equations with drift-admitting jumps. To satisfy the matching conditions at the jumps, i.e., the continuities of the propagator and the probability current, the jumps are set to be solution points and used to divide the solution domain into subdomains. While the values of the probability current at flux points are obtained within each subdomain, the values of its first derivative at solution points are evaluated by using stencils across the subdomains. Several benchmark problems are solved numerically to show the validity of the proposed scheme.
\end{abstract}

\pacs{}

\maketitle

\section{Introduction}

Piecewise-smooth stochastic systems are used as models of physical and biological systems \cite{Reimann2002,Gennes2005dryfriction,SanoKanazawa2016}. The interrelation between noise and discontinuities in such systems has attracted considerable attention recently
 \cite{KawaradaHayakawa2004Non-Gaussian,Hayakawa2005Langevin,BauleCohenTouchette2010path,MenzelGoldenfeld2011,BauleTouchetteCohen2011path,
BauleSollich2012,BauleSollich2013,ChenJust2013,ChenJust2014,SanoHayakawa2014,GeffertJust2017}.
Some of them can be modeled by stochastic differential equations (SDEs) with piecewise-smooth drifts. Particularly, we consider in this paper the problems that can be modeled by the Langevin equation
\begin{equation}
   \dot{v}(t)=\Phi(v)+\sqrt{2D}\xi(t),
   \label{aa}
\end{equation}
where the overdot denotes the time derivative, the drift $\Phi(v)$ is discontinuous at some points, and $D>0$ represents the strength of the Gaussian white noise $\xi(t)$ that is characterized by the zero mean $\langle \xi(t) \rangle=0$ and the correlation $\langle \xi(t)\xi(t') \rangle=\delta(t-t')$. Here, the notation $\langle \cdots \rangle$ stands for the average over all possible realizations of the noise, and $\delta$ denotes the Dirac delta function. The initial condition for Eq.~(\ref{aa}) is set to be $v(0)=v_0$.

The theory of piecewise-smooth SDEs is only in its infancy
compared to its noiseless counterpart \cite{BernardoBudd2008}. For a few simple piecewise-smooth drifts, the propagators of Eq.~(\ref{aa}) are known analytically. For instance, when the drift is pure dry friction \cite{Gennes2005dryfriction} (also called solid friction or Coulomb friction), the
propagator is available in closed analytic form \cite{CaugheyDienes1961,Karatzas1984,TouchetteStraeten2010Brownian}. More generally, when the
drift is piecewise constant with a discontinuity (called the Brownian motion with a two-valued drift), the propagator can be expressed in terms of convolution integrals \cite{Karatzas1984,SimpsonKuske2014TwoValued}. Moreover, the distribution of the occupation time can also be  obtained analytically \cite{Simpson2014OPT}. When the drift contains both dry friction and viscous friction, the  propagator can be expressed as a sum of series \cite{TouchetteStraeten2010Brownian} or in connection with a Laplace transform \cite{TouchetteThomas2012Brownian}. For Eq.~(\ref{aa}) with dry friction the first two moments of the displacement and other integral functionals have also been obtained by solving backward Komogorov equations \cite{ChenJust2014II} or using the method based on the Pugachev-Sveshnikov equation \cite{Berezin2018}. However, there are vast cases that cannot be solved analytically by using existing theoretical methods. In those cases, we should resort to some effective numerical methods if we want to know the dynamics of Eq.~(\ref{aa}).

For instance, one can employ some numerical schemes to solve the SDE (\ref{aa}) directly. The Euler-Maruyama scheme is one of the simplest schemes that can be applied to obtain approximate results \cite{Leobacher2016}. However, there are errors arising from the approximations to discontinuities and the derivative. To address this issue, the so-called exact simulation was developed for solving Brownian motions with drift admitting a unique jump \cite{Etore2014, papaspiliopoulos2016}. The exact simulation involves only computer representation errors, enabling one to get exact samplings for the considered SDEs. In addition, the algorithm can be generalized to solve Brownian motions with drift admitting several jumps \cite{Dereudre2017}. Nevertheless, it requires heavy calculations to realize the exact simulation.

In this paper, we intend to solve the following Fokker-Planck equation directly, which governs the propagator of the model (\ref{aa}) with the Gaussian white noise:
\begin{equation}
   \partial_t\, p=-\partial_v[\Phi(v)p]+D\partial^2_v p,
\label{ac}
\end{equation}
where $p=p(v,t|v_0,0)$ denotes the propagator with the initial condition  $p(v,0|v_0,0)=\delta(v-v_0)$.
To solve Eq.~(\ref{ac}) with drift-admitting jumps, we need to apply two matching conditions at each jump of the drift, i.e., the continuity of the propagator and the continuity of the probability current (or flux)
\begin{equation}
  f(v,t|v_0,0)=-\Phi(v)p+D\partial_v\, p. \label{ad}
\end{equation}
When the drift is continuous, there are many numerical methods that can be used to solve Eq.~(\ref{ac}); see for instance \cite{ChangCooper1970,LarsenLevermore1985,Langtangen1991,DrozdovMorillo1996,ZhangWei1997,Wei2000,LiuYu2014,PareschiZanella2018}. However, to the best of our knowledge, there are only few numerical results in the literature considering the cases with drift-admitting jumps. In \cite{MenzelGoldenfeld2011}, the authors transformed the Fokker-Planck equation with pure dry friction to a Schr\"{o}dinger equation with a delta potential, and then investigated the displacement statistics by solving a corresponding Brinkman hierarchy numerically. By treating the discontinuous drift carefully using a finite volume method \cite{ZhangChen2018} or an immerse interface method \cite{ZhangChen2017}, second-order schemes were developed for solving Eq.~(\ref{ac}). In this paper, we attempt to derive a finite difference scheme based on a grid staggered by flux points and solution points (see e.g. Fig.~\ref{fig_grid}). It will be seen later that the aforementioned matching conditions at jumps can be easily satisfied by using this grid, resulting in a simple way to treat the cases with drift-admitting jumps.

The rest of this paper is arranged as follows. In Sec.~\ref{sec_2} we take as an example the case with drift admitting two jumps to describe the procedure of the main algorithm for the \blue{spatial} discretization. The corresponding staggered grid is also introduced. Then we present the finite difference scheme in Sec.~\ref{sec_3}. Some benchmark problems are solved numerically in Sec.~\ref{sec_4} to show the validity of the scheme. In Sec.~\ref{sec_5}, we extend the algorithm to study the displacement of the Brownian motion with pure dry friction. Finally, conclusions are drawn in Sec.~\ref{sec_6}.

\section{Staggered grid}
\label{sec_2}

We describe the algorithm by assuming that the drift in Eq.~(\ref{aa}) admits two jumps at $v=v_{d_1}, v_{d_2}$  ($v_{d_1}<v_{d_2}$)\red{, respectively}. For other cases, the algorithm can be generalized straightforwardly
according to the number of jumps.

\begin{figure*}
  \begin{center}
   \includegraphics[width=0.9\linewidth]{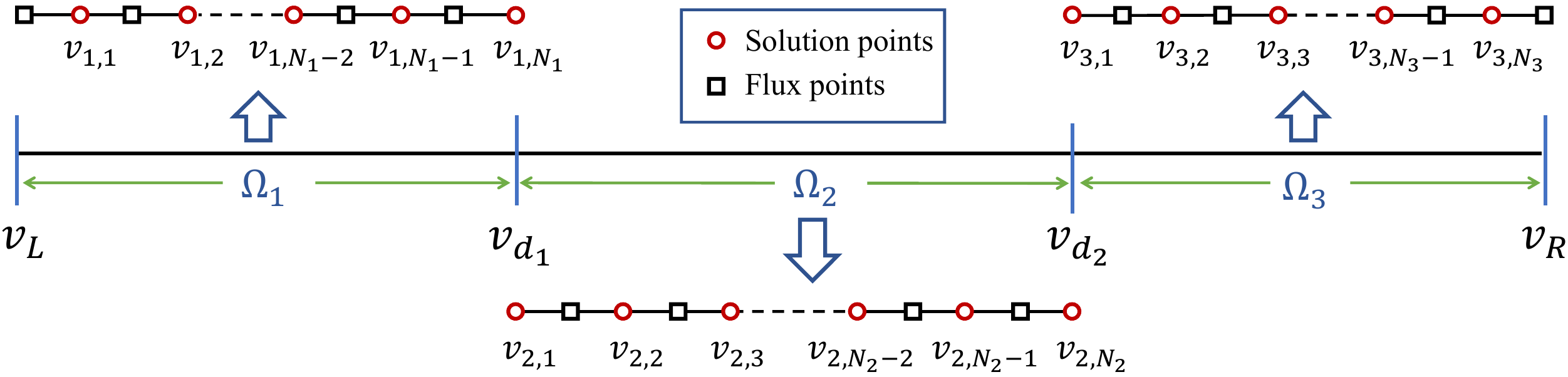}
   \end{center}
   \caption{Illustration of the grid \blue{staggered by flux points and solution points for the case with two jumps at $v=v_{d_1}$ and $v=v_{d_2}$, respectively. The two jumps are both set to be
   solution points and used to divide the computational domain $[v_{_L},v_{_R}]$ into three subdomains $\Omega_i$ ($i=1,2,3$).}}
   \label{fig_grid}
\end{figure*}

For Eq.~(\ref{ac}) defined for $v\in (-\infty,\infty)$, we first truncate the domain into a finite interval, denoted by $[v_{_L},v_{_R}]$, containing the two discontinuous points. Then by using these two points we partition the interval into three subdomains, i.e., $\Omega_1=[v_{_L}, v_{d_1}]$, $\Omega_2=[v_{d_1}, v_{d_2}]$ and $\Omega_3=[v_{d_2},v_{_R}]$. As illustrated in Fig.~\ref{fig_grid}, a grid staggered by flux points and solution points is used for the partitioned subdomains. In particular, the discontinuous points $v_{d_1}$ and $v_{d_2}$ are both set to be solution points such that the continuity condition\red{s} of the propagator are satisfied automatically for the discrete method.

In each subdomain $\Omega_i$, the grid points are set to be uniformly distributed with the solution points defined by
\begin{align}
   \begin{cases}
   v_{1,j}=v_{_L}+(j-1/2)h_1,  & 1\leqslant j \leqslant N_1,\\
   v_{2,j}=v_{d_1}+(j-1)h_2,   & 1\leqslant j \leqslant N_2,\\
   v_{3,j}=v_{d_2}+(j-1)h_3, & 1\leqslant j \leqslant N_3,\\
   \end{cases}
\end{align}
where $N_i$ are the numbers of solution points and $h_i$ the \blue{spatial} steps for the subdomains,
\begin{align}
   \begin{cases}
   h_1=(v_{d_1}-v_{_L})/(N_1-1/2),\\
   h_2=(v_{d_2}-v_{d_1})/(N_2-1),\\
   h_3=(v_{_R}-v_{d_2})/(N_3-1/2).
   \end{cases}
\end{align}
Especially, we have $
   v_{1,N_1}=v_{2,1}=v_{d_1}
$ and
$
   v_{2,N_2}=v_{3,1}=v_{d_2}
$.
The flux points $v_{i,j+1/2}$ are defined by
\begin{equation}
   \begin{cases}
   v_{1,j+1/2}=v_{_L}+jh_1,  & 0\leqslant j \leqslant N_1-1,\\
   v_{2,j+1/2}=v_{d_1}+(j-1/2)h_2,   & 1\leqslant j \leqslant N_2-1,\\
   v_{3,j+1/2}=v_{d_2}+(j-1/2)h_3, & 1\leqslant j \leqslant N_3.\\
   \end{cases}
\end{equation}
Particularly, we have $ v_{1,1/2}=v_{_L} $ and $ v_{3,N_3+1/2} = v_{_R} $, i.e., the end points
of the interval $ [v_{_L}, v_{_R}] $ are both flux points, which are designed to impose boundary conditions.

\begin{figure}
   \begin{center}
      \includegraphics[width=0.75\linewidth]{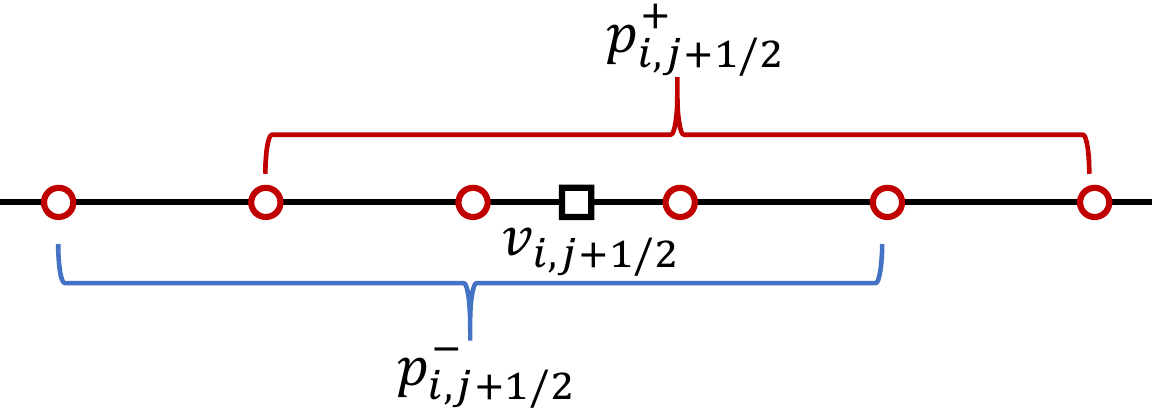}
   \end{center}
   \caption{Illustration of the stencils \blue{ used to reconstruct the values $p_{i,j+1/2}^\pm$ required by  the approximation (\ref{eq_drift_spliting}). Here only the stencils of the fifth-order interior interpolations are presented. Near the boundaries the stencils should be adjusted accordingly. } }
   \label{fig_grid_ill}
\end{figure}

Given initial values at the solution points of the staggered grid, a finite difference scheme \red{for Eq.~(\ref{ac})} can be constructed by the following procedure:
\begin{enumerate}
   \item[(i)] Within each subdomain $\Omega_i$, obtain the solutions at the flux points by using interpolation schemes. For the purpose of stability, \emph{upwind interpolations} are used here according to the sign of the drift $\Phi(v)$. If $\Phi(v)$ changes its sign within $\Omega_i$, \mage{we need to
       split the drift into an appropriate form to apply upwind interpolations. Here we split the drift into two parts: $\Phi(v)=\min\{\Phi(v),0\}+\max\{\Phi(v),0 \}$, ensuring that each part does not change its sign.       Then using this split form we can approximate the term $\Phi(v)p$ appearing in Eq.~(\ref{ac}) at the flux point $v_{i,j+1/2}$ by}
       \begin{align}
          \min&\{\Phi(v_{i,j+1/2}),0\}p^+_{i,j+1/2}\nonumber\\
          &+\max\{\Phi(v_{i,j+1/2}),0 \}p^-_{i,j+1/2},
          \label{eq_drift_spliting}
       \end{align}
       where $p^+_{i,j+1/2}$ and $p^-_{i,j+1/2}$ are the approximate values of $p$ at $v_{i,j+1/2}$, respectively, obtained by using interpolations with
       stencils as illustrated in Fig.~\ref{fig_grid_ill};
   \item[(ii)] Evaluate the first derivative of $p$ at flux points by using
         \emph{difference schemes} in each subdomain $\Omega_i$;
   \item[(iii)] Obtain the values of the current (\ref{ad}) at fluxes points by using the above two steps. Then approximate the values of the derivative of the current at solution points by using a difference scheme, which is designed for the domain $[v_{_L}, v_{_R}]$ directly since the current (\ref{ad}) is theoretically continuous everywhere.
\end{enumerate}

\section{Scheme}
\label{sec_3}

To discretize the right side of Eq.~(\ref{ac}), we follow the aforementioned procedure: first calculate the values of the probability current at flux points and then derive a difference scheme to evaluate the derivative of the current. Here the \blue{spatial} scheme is designed to be fifth-order for the cases with smooth drifts. Finally, a third-order Runge-Kutta scheme is employed to solve the resulting ordinary differential system.

\subsection{Evaluation of the \red{p}robability current}
\label{sec_3a}

There are two terms appearing in the current (\ref{ad}). For the first term $\Phi(v)p$, we use interpolation schemes to reconstruct the required values in the \red{approximation} (\ref{eq_drift_spliting}). For the second term $D\partial_v p$, we derive difference schemes to approximate it.

\blue{In the following, we will present the schemes in matrix forms, where the entries of the matrices are all easily obtained by using Lagrangian interpolations according to specified stencils.
For example, if we consider the stencil $S=\{v_{1,1},v_{1,2},\dots,v_{1,5}\}$,
then at any point $v$, the values of $p$ and $\partial_v p$ are approximated respectively by
\begin{align}
   p_{_I}(v)=&\sum_{k=1}^5 l_k(v) p_{1,k}, \\
   \frac{d}{dv} p_{_I}(v)=&\sum_{k=1}^5 \frac{d}{dv} l_k(v) p_{1,k},
\end{align}
where
\begin{equation}
   l_{k}(v)=\prod_{s=1,s\neq k}^5 \frac{v-v_{1,s}}{v_{1,k}-v_{1,s}}.
\end{equation}
}

Since the grid points are slightly different in different subdomains (see Fig.~\ref{fig_grid}),
we describe the schemes separately for each subdomain $\Omega_i$.
\blue{For convenience of notations, we will first consider the subdomain $\Omega_2$, and then
$\Omega_1$ and $\Omega_3$.}

\subsubsection{Subdomain $\Omega_2$}

To describe the scheme compactly, let us introduce the vectors
   $\mathbf{p}^\pm_2=[ p^\pm_{2,3/2},p^\pm_{2,5/2},\dots,p^\pm_{2,N_2-1/2} ]^T$ and
  the vector $\mathbf{p}_2=[ p_{2,1},p_{2,2},\dots,p_{2,N_2} ]^T$,
where the ``\blue{$T$}'' denotes the transpose operation. Then according to the grid point distribution in $\Omega_2$ we can compute the values at flux points by the fifth-order interpolation schemes $ \mathbf{p}^\pm_2=I_2^\pm \mathbf{p}_2 $, \blue{where $I_2^\pm$ are both $(N_2-1) \times N_2$
matrices. Here}
\begin{equation}
I_2^+=\begin{bmatrix}
\frac{35}{128} & \frac{35}{32} & -\frac{35}{64} & \frac{7}{32} &
   -\frac{5}{128} &  &  \\[3pt]
 -\frac{5}{128} & \frac{15}{32} & \frac{45}{64} & -\frac{5}{32} &
   \frac{3}{128} &  &  \\[3pt]
  & \ddots & \ddots & \ddots & \ddots &
   \ddots &  \\[3pt]
  &  & -\frac{5}{128} & \frac{15}{32} & \frac{45}{64} & -\frac{5}{32} &
   \frac{3}{128} \\[3pt]
  &  & \frac{3}{128} & -\frac{5}{32} & \frac{45}{64} & \frac{15}{32} &
   -\frac{5}{128} \\[3pt]
  &  & -\frac{5}{128} & \frac{7}{32} & -\frac{35}{64} & \frac{35}{32} &
   \frac{35}{128}
\end{bmatrix},
  \label{eq_matrix_I2p}
\end{equation}
and $I_2^-$ is defined by letting its entries satisfy that $[I_2^-]_{j,k}=[I_2^+]_{N_2-j,N_2+1-k}$, $1\leqslant j \leqslant N_2-1$, $1\leqslant k \leqslant N_2$.

Denote the approximation to $\partial_v p$ at the flux point $v_{2,j+1/2}$ as $(\partial_v p)_{2,j+1/2}$ and introduce the vector
 \[\mathbf{p}_{2,v}=[ (\partial_v p)_{2,3/2},(\partial_v p)_{2,5/2},\dots,(\partial_v p)_{2,N_2-1/2}]^T.\]
We can write the difference scheme as
$
  \mathbf{p}_{2,v}=A_2\mathbf{p}_2/h_2,
$
where the $(N_2-1)\times N_2$ matrix $A_2$ is
\begin{equation}
A_2=\begin{bmatrix}
 -\frac{11}{12} & \frac{17}{24} & \frac{3}{8} & -\frac{5}{24} & \frac{1}{24} &
   &   &  \\[3pt]
 \frac{1}{24} & -\frac{9}{8} & \frac{9}{8} & -\frac{1}{24} &  &  &  &  \\[3pt]
 -\frac{3}{640} & \frac{25}{384} & -\frac{75}{64} & \frac{75}{64} &
   -\frac{25}{384} & \frac{3}{640} &  &  \\[3pt]
 &  \ddots & \ddots & \ddots & \ddots &
   \ddots & \ddots &  \\[3pt]
 &   & -\frac{3}{640} & \frac{25}{384} & -\frac{75}{64} & \frac{75}{64} &
   -\frac{25}{384} & \frac{3}{640} \\[3pt]
 &   &  &  & \frac{1}{24} & -\frac{9}{8} & \frac{9}{8} & -\frac{1}{24} \\[3pt]
 &   &  & -\frac{1}{24} & \frac{5}{24} & -\frac{3}{8} & -\frac{17}{24} &
   \frac{11}{12}
\end{bmatrix},
  \label{eq_matrix_a2}
\end{equation}
such that the difference scheme is sixth-order at $v_{2,j+1/2}$ with $3\leqslant j \leqslant N_2-3$ and fourth-order at the other flux points.

\subsubsection{Subdomain $\Omega_1$}
Introduce $\mathbf{p}^\pm_1=[ p^\pm_{1,1/2},p^\pm_{1,3/2},\dots,p^\pm_{1,N_1-1/2} ]^T$ and $\mathbf{p}_1=[ p_{1,1},p_{1,2},\dots,p_{1,N_1} ]^T$.
 We first compute the right vector $\mathbf{p}^+_1$ by using the fifth-order interpolation scheme $ \mathbf{p}^+_1=I_1^+\mathbf{p}_1 $ with the $N_1\times N_1$ matrix $I_1^+$ written as
\blue{
\begin{equation*}
I_1^+=\begin{bmatrix}
  \mathbf{a} \\
  (I_2^+)_{(N_1-1)\times N_1}
\end{bmatrix}.
\end{equation*}
Here
\begin{equation}
 \mathbf{a}=
\begin{bmatrix}
     \frac{315}{128} & -\frac{105}{32} & \frac{189}{64} & -\frac{45}{32} &
   \frac{35}{128} & 0 & \dots & 0
 \end{bmatrix}
 \label{eq_def_b}
\end{equation}
is a $1\times N_1$ vector and the matrix $(I_2^+)_{(N_1-1)\times N_1}$ is defined in Eq.~(\ref{eq_matrix_I2p}) by replacing $N_2$ with $N_1$ (The same notation method will be used throughout this paper).
}For the left values, first let $p^-_{1,1/2}=p^+_{1,1/2}$. Then we derive the interpolation scheme $ \mathbf{p}^-_1=I_1^-[p^-_{1,1/2},\mathbf{p}_1^T]^T $ with the $N_1\times (N_1+1)$
matrix $I_1^-$ reading as
\begin{equation}
  I_1^-= \begin{bmatrix}
1 &  &  &  &  &  &  &  \\[3pt]
 -\frac{1}{7} & \frac{5}{8} & \frac{5}{8} & -\frac{1}{8} & \frac{1}{56} &  &
    &  \\[3pt]
 \frac{3}{35} & -\frac{1}{4} & \frac{3}{4} & \frac{9}{20} & -\frac{1}{28} &
   &  &  \\[3pt]
  & \frac{3}{128} & -\frac{5}{32} & \frac{45}{64} & \frac{15}{32} &
   -\frac{5}{128} &  &  \\[3pt]
  &  & \ddots & \ddots & \ddots & \ddots &
   \ddots &  \\[3pt]
  &  &  & \frac{3}{128} & -\frac{5}{32} & \frac{45}{64} & \frac{15}{32} &
   -\frac{5}{128} \\[3pt]
  &  &  & -\frac{5}{128} & \frac{7}{32} & -\frac{35}{64} & \frac{35}{32} &
   \frac{35}{128}
   \end{bmatrix}.
   \label{eq_matrix_I1m}
\end{equation}

Introducing the vector
 \[\mathbf{p}_{1,v}=[ (\partial_v p)_{1,1/2},(\partial_v p)_{1,3/2},\dots,(\partial_v p)_{1,N_1-1/2} ]^T,\]
we can derive the difference scheme $ \mathbf{p}_{1,v}=A_1 \mathbf{p}_1/h_1 $ with the $N_1\times N_1$ matrix $A_1$ written as
\blue{
\begin{equation*}
  A_1=\begin{bmatrix}
    \mathbf{b}\\
 (A_2)_{(N_1-1)\times N_1}
   \end{bmatrix},
\end{equation*}
where
\begin{equation}
   \mathbf{b}=\begin{bmatrix}
-\frac{31}{8} & \frac{229}{24} & -\frac{75}{8} & \frac{37}{8} &
   -\frac{11}{12} & 0 & \dots & 0 \\
   \end{bmatrix}
\end{equation}
is a $1\times N_1$ vector and $A_2$ is defined in Eq.~(\ref{eq_matrix_a2}),
such that } the difference scheme is sixth-order at $v_{1,j+1/2}$ with $ 3\leqslant j\leqslant N_1-3$ and fourth-order at the other flux points.

\subsubsection{Subdomain $\Omega_3$}

\blue{The schemes for this subdomain are basically the same as
those of the subdomain $\Omega_2$ if we swap the direction.} Introduce
$\mathbf{p}^\pm_3=[ p^\pm_{3,3/2},p^\pm_{3,5/2},\dots,p^\pm_{3,N_3+1/2} ]^T$ and  $\mathbf{p}_3=[ p_{3,1},p_{3,2},\dots,p_{3,N_3} ]^T$.
We first reconstruct the vector $\mathbf{p}^-_3$ by using the fifth-order interpolation scheme $\mathbf{p}^-_3=I_3^-\mathbf{p}_3$ with \blue{the entries of} the $N_3\times N_3$ matrix $I_3^-$ defined by
\blue{
$
[I_3^-]_{j,k}=[(I_1^+)_{N_3\times N_3}]_{N_3+1-j,N_3+1-k}
$.
}Then the right values are calculated by the fifth-order interpolation scheme $\mathbf{p}^+_3=I_3^+[\mathbf{p}_3^T,p^+_{3,N_3-1/2}]^T$, where the
entries of the $N_3\times (N_3+1)$ matrix $I_3^+$ are derived to be
\blue{
$
[I_3^+]_{j,k}=[(I_1^-)_{N_3\times (N_3+1)}]_{N_3+1-j,N_3+2-k}
$}. Note that we have assumed $p^+_{3,N_3-1/2}=p^-_{3,N_3-1/2}$ here.

Similarly, by introducing the vector
 \[
 \mathbf{p}_{3,v}=[ (\partial_v p)_{3,3/2},(\partial_v p)_{3,5/2},\dots,(\partial_v p)_{3,N_3+1/2} ]^T,
 \]
 the derivative values at flux points are approximated by
 $ \mathbf{p}_{3,v}=A_3 \mathbf{p}_{3} /h_3 $ with \blue{the entries} of the $N_3\times N_3$ matrix $A_3$ defined by  $\blue{ [A_3]_{j,k}=-[(A_1)_{N_3\times N_3}]_{N_3+1-j,N_3+1-k}}$, such that the difference scheme is sixth-order at $v_{3,j+1/2}$ with $3\leqslant j \leqslant N_3-3$ and fourth-order at the other flux points.

\subsubsection{Imposing boundary conditions}

Using the above schemes derived for subdomains, we can obtain the values of the probability current at all flux points. However, it is noted that we have not used any boundary conditions
so far. As we will see later in Sec.~\ref{sec_4}, depending on the signs of the drift at the domain boundaries we may need to set the computational domain to be large enough and impose reflecting boundary conditions \cite{Veestraeten2004} appropriately. In that cases, we just \red{reset} the current values at the boundaries to be zero.

\subsection{Derivative of the current}

Now we are ready to derive a difference scheme to compute the derivative of the current using
the values at flux points obtained in Sec.~\ref{sec_3a}. To get a correct solution, it is no doubt that we have to consider information transmission between different subdomains $\Omega_i$. As mentioned before, although the derivative of the propagator is not continuous at jumps, the current is continuous everywhere. Therefore, we can derive a difference scheme for the whole domain directly to approximate the derivative of the current. However, as it allows different \blue{spatial} steps in different subdomains, we have to pay attention to the solution points near the jumps.

For convenience of notations, let us introduce the flux vector
$
   \mathbf{f}=[ \mathbf{f}_1^T,\mathbf{f}_2^T,\mathbf{f}_3^T ]^T
$, where
\begin{align}
   \mathbf{f}_1=[ f_{1,1/2},f_{1,3/2},\dots, f_{1,N_1-1/2} ]^T,\\
   \mathbf{f}_2=[ f_{2,3/2},f_{1,5/2},\dots,f_{2,N_2-1/2} ]^T,\\
   \mathbf{f}_3=[ f_{3,3/2},f_{1,5/2},\dots,f_{3,N_3+1/2} ]^T.
\end{align}
The values of the derivative $\partial_v f$ at solution points are denoted by
$
   \mathbf{f}_v=[ \mathbf{f}_{1,v}^T,\mathbf{f}_{2,v}^T,\mathbf{f}_{3,v}^{T} ]^T
$ with
\begin{align}
   \mathbf{f}_{1,v}=[ (\partial_v f)_{1,1},(\partial_v f)_{1,2},\dots,(\partial_v f)_{1,N_1} ]^T, \\
   \mathbf{f}_{2,v}=[ (\partial_v f)_{2,2},(\partial_v f)_{2,3},\dots,(\partial_v f)_{2,N_2} ]^T, \\
   \mathbf{f}_{3,v}=[ (\partial_v f)_{3,2},(\partial_v f)_{3,3},\dots,(\partial_v f)_{3,N_3} ]^T.
\end{align}
Then we attempt to derive a \red{derivative} matrix $A$ such that
$\mathbf{f}_v=A \mathbf{f}$. Here the size of $A$ is $N_v \times (N_v+1)$ with $N_v=N_1+N_2+N_3-2$.

By observing the distribution of the grid points, we design the difference scheme by
using the stencils
\begin{align}
   [\mathbf{f}_v]_j =
   \begin{cases}
      \sum_{k=1}^5 a_{j,k}[\mathbf{f}]_k, & 1\leqslant j\leqslant 2, \\[2pt]
      \sum_{k=1}^6 a_{j,k}[\mathbf{f}]_{j+k-3}, & 3\leqslant j \leqslant N_v-2,\\[2pt]
      \sum_{k=1}^5 a_{j,k}[\mathbf{f}]_{N_v+k-4}, & N_v-1\leqslant j \leqslant N_v,\\
   \end{cases}
   \label{eq_diff}
\end{align}
where $[\mathbf{f}]_k$ denotes the $k$-th entry of the vector $\mathbf{f}$ and
the coefficients $a_{j,k}$ can be determined directly by using Lagrangian interpolations.
Hence the difference matrix $A$ can be easily written down following the above stencils.

\begin{figure}
   \begin{center}
      \includegraphics[width=0.8\linewidth]{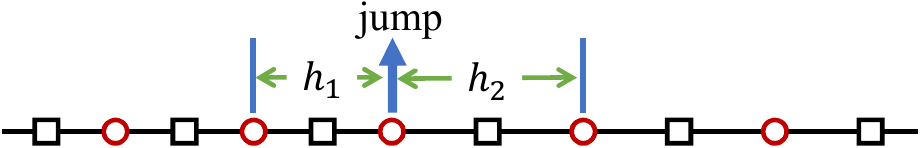}
   \end{center}
   \caption{Illustration of the grid points near a jump. \blue{ The spatial steps on the left and on the right of the jump are $h_1$ and $h_2$, respectively.} }
   \label{fig_discontinuities}
\end{figure}

We first present the coefficients of the cases with stencils in a single subdomain.
The results are as follows:
\begin{align}
[a_{j,k}]_{1\leqslant k \leqslant 5}&=
   \begin{cases}
      \frac{1}{h_1}
[
 -\frac{11}{12} , \frac{17}{24} , \frac{3}{8} , -\frac{5}{24} , \frac{1}{24}
], & j=1, \\[2pt]
\frac{1}{h_1}
[
 \frac{1}{24} , -\frac{9}{8} , \frac{9}{8} , -\frac{1}{24} , 0
], & j=2,
   \end{cases}
\\
[a_{j,k}]_{1\leqslant k \leqslant 6}&=
   \begin{cases}
      \frac{1}{h_1} \mathbf{c}, & 3\leqslant j\leqslant N_1-3,\\[2pt]
      \frac{1}{h_2} \mathbf{c}, & N_1+3\leqslant j \leqslant N_1+N_2-4,\\[2pt]
      \frac{1}{h_3} \mathbf{c}, & N_1+N_2+2\leqslant j \leqslant N_v-2,
   \end{cases}
\\
[a_{j,k}]_{1\leqslant k \leqslant 5}&=
   \begin{cases}
  \frac{1}{h_3}
 [
 0 , \frac{1}{24} , -\frac{9}{8} , \frac{9}{8} , -\frac{1}{24}
 ], &  j=N_v-1, \\[2pt]
\frac{1}{h_3}[
 -\frac{1}{24}, \frac{5}{24}, -\frac{3}{8}, -\frac{17}{24}, \frac{11}{12}], & j=N_v,
   \end{cases}
\end{align}
where the vector
\begin{equation}
\mathbf{c} =
   \begin{bmatrix}
   -\tfrac{3}{640} & \tfrac{25}{384} & -\tfrac{75}{64} &
 \tfrac{75}{64} & -\tfrac{25}{384} & \tfrac{3}{640}
   \end{bmatrix}.
\end{equation}
For the other cases, we have to pay attention to the fact that the stencils (\ref{eq_diff}) are across the jumps, as illustrated in Fig.~\ref{fig_discontinuities}.
When $N_1-2\leqslant j \leqslant N_1+2$, \red{by using Lagrangian interpolations}
we can determine the coefficients to be
\begin{equation}
   a_{j,k}=d_{j-N_1+3,k}(h_1,h_2), \quad 1\leqslant k \leqslant 6,
\end{equation}
where $d_{s,k}(x,y)$ are functions of $x$ and $y$, as shown in Tab.~\ref{diff_coefficients_cross}.
Similarly, when $N_1+N_2-3\leqslant j \leqslant N_1+N_2+1$, we have
\begin{equation}
   a_{j,k}=d_{j-N_1-N_2+4,k}(h_2,h_3), \quad 1\leqslant k \leqslant 6.
\end{equation}

   \begin{table*}
     \begin{center}
   \begin{tabular*}{\linewidth}{@{\extracolsep{\fill}}cccc}
      \hline
     & $ s=1 $ & $ s=2 $ & $ s=3 $ \\
     \cline{2-4}
     \vspace{-0.8em}\\
     $d_{s,1}(x,y)$
     & $ -\frac{3}{64 (9 x+y)} $
     & $ \frac{-8 x^2-4 x y+3 y^2}{24 x (7 x+y) (7 x+3 y)} $
     & $ \frac{60 y^3-69 x y^2}{20 x(x+y) (5 x+y) (5 x+3 y)} $
     \\[4pt]
     $d_{s,2}(x,y)$
     & $ \frac{23 x+2 y}{336 x^2+48 x y}$
     & $ \frac{16 x^2+12 x y-3 y^2}{8 x (5 x+y) (5 x+3 y)} $
     & $ \frac{5 (23 x-18 y) y^2}{6 x (x+y) (3 x+y) (3 x+5 y)} $
     \\[4pt]
     $d_{s,3}(x,y)$
     & $ -\frac{9 (21 x+4 y)}{32 x (5 x+y)}$
     & $ -\frac{88 x^2+116 x y+21 y^2}{24 x (x+y) (3 x+y)} $
     & $ \frac{15 y^2 (8 y-23 x)}{4 x (x+y) (x+3 y) (x+5 y)} $
     \\[4pt]
     $d_{s,4}(x,y)$
     & $ \frac{57 x+18 y}{48 x^2+16 x y} $
     & $ \frac{32 x^2+124 x y+69 y^2}{24 x (x+y) (x+3 y)} $
     & $ \frac{15 x^2 (23 y-8 x)}{4 y (x+y) (3 x+y) (5 x+y)} $
     \\[4pt]
     $d_{s,5}(x,y)$
     & $ -\frac{17 x+8 y}{192 x (x+y)} $
     & $ -\frac{x^3 (x+24 y)}{y (x+y) (3 x+y) (5 x+y) (7 x+y)} $
     & $ \frac{5 x^2 (18 x-23 y)}{6 y (x+y) (x+3 y) (5 x+3 y)} $
     \\[4pt]
     $d_{s,6}(x,y)$
     & $ \frac{18 x^4}{(x+y) (3 x+y) (5 x+y) (7 x+y) (9 x+y)} $
     & $ \frac{x^3 (x+8 y)}{3 y (x+y) (x+3 y) (5 x+3 y) (7 x+3 y)} $
     & $ \frac{3 x^2 (23 y-20 x)}{20 y (x+y) (x+5 y) (3 x+5 y)} $ \\[4pt]
     \hline
   \end{tabular*}
     \end{center}
   \caption{Coefficients that determine the difference scheme (\ref{eq_diff}) for the cases $N_1-2\leqslant j \leqslant N_1+2$ and $N_1+N_2-3\leqslant j \leqslant N_1+N_2+1$. \blue{When
   $s=4$ and $5$, the coefficients are determined by the relation $d_{s,j}(x,y)=-d_{6-s,7-j}(y,x)$ for
   $1\leqslant j \leqslant 6$.}}
   \label{diff_coefficients_cross}
   \end{table*}

\subsection{Time-marching scheme}

Approximating the right side of Eq.~(\ref{ac})
by using the above finite difference scheme, we obtain
a semi-discretized system, denoted by
\begin{equation}
  \frac{d \mathbf{p}}{dt}= R(\mathbf{p},t),
\end{equation}
where $\mathbf{p}$ stands for the vector of the unknowns at solution points, and $R(\mathbf{p},t)$
represents the right hand side term.  Then many time-marching schemes can be used to solve this
system. In this paper we employ a traditional third-order Runge-Kutta scheme, written as
\begin{align}
   \mathbf{p}^{n+1}=\mathbf{p}^n+\frac{1}{9}\tau (2K_1+3K_2+4K_3),\label{rk_1}\\
   \begin{cases}
      K_1=R( \mathbf{p}^n, t_n ),\\
      K_2=R( \mathbf{p}^n+\frac{1}{2}\tau K_1, t_n+\frac{1}{2}\tau ),\\
      K_3=R( \mathbf{p}^n+\frac{3}{4}\tau K_2, t_n+\frac{3}{4}\tau ),
   \end{cases}
   \label{rk_2}
\end{align}
where $\mathbf{p}^n$ denotes the value of $\mathbf{p}$ at time $t_n$ and $\tau$ is the time step.

\section{Numerical examples}
\label{sec_4}

In this section, some benchmark problems are solved numerically to  show
the validity of the algorithm presented above. The discrete $L^2$-norm error for the case with two jumps is defined by
\begin{align}
   L^2 \mbox{ error}=&\Bigg[\sum_{j=1}^{N_1-1} e_{1,j}^2h_1 +\frac{1}{2}e_{1,N_1}^2(h_1+h_2)+\sum_{j=2}^{N_2-1}e_{2,j}^2h_2\nonumber\\
   &+\frac{1}{2} e_{2,N_2}^2(h_2+h_3)+\sum_{j=2}^{N_3}e_{3,j}^2h_3\Bigg]^{1/2},
   \label{eq_l2error}
\end{align}
where $e_{i,j}=p_{i,j}-p(v_{i,j})$ are the errors between numerical results and exact solutions.
\red{In addition,} the $L^\infty$-norm error is defined by
\begin{equation}
  L^\infty \mbox{ error}=\max_{i,j}\{|e_{i,j}|\}.
  \label{eq_linfty}
\end{equation}
For other cases, the errors are defined similarly.
The numerical convergence rate is defined by
\begin{equation}
     \mbox{rate}=-\ln( E_{M}/E_{N} )/\ln( M/N ),
     \label{eq_rate}
\end{equation}
where $E_{M}$ and $E_{N}$ are the errors corresponding to the cases with $M$ and $N$ solution points, respectively.

In numerical computations, the initial condition of Eq.~(\ref{ac}) given by a delta function cannot be used directly. Instead, if an exact solution to Eq.~(\ref{ac}) is available we choose the initial condition to be $p(v,\tau_0|v_0,0)$ and start the computation from $t=\tau_0$. Here
$\tau_0$ is a constant that can be chosen appropriately for the considered problems. Otherwise,
the initial condition is set to be Gaussian,
\begin{equation}
      p(v,\tau_0|v_0,0)=\frac{1}{ \sqrt{4\pi D \tau_0}  }e^{ -[v-v_0-\Phi(v_0)\tau_0]^2/(4D \tau_0) },
      \label{eq_v}
\end{equation}
which mimics the delta function when $\tau_0$ is small. For convenience, \red{$D=0.5$ and $\tau_0=0.01$} are chosen for all test cases in this section.

It should be noted that the proposed finite difference scheme can also be applied to solve problems with continuous drifts. In the following, we first show that the scheme is actually fifth-order for smooth cases. Then we pay attention to the cases with drift-admitting jumps, where a second-order convergence rate is observed.

\subsection{Smooth drifts}

The following \red{two} examples are used to confirm \red{that
the scheme described in Sec.~\ref{sec_3} is fifth-order for the cases with smooth drifts.}
Here $v_{d_1}=0$ and $v_{d_2}=1$ are used to divide all the computational domains (see Fig.~\ref{fig_grid}) and the time step is set to be $\tau=0.01\min\{h_1^2,h_2^2,h_3^2\}$.

\subsubsection{Constant drift} \label{sec_4aa}
\begin{figure}
   \begin{center}
       \includegraphics[width=0.9\linewidth]{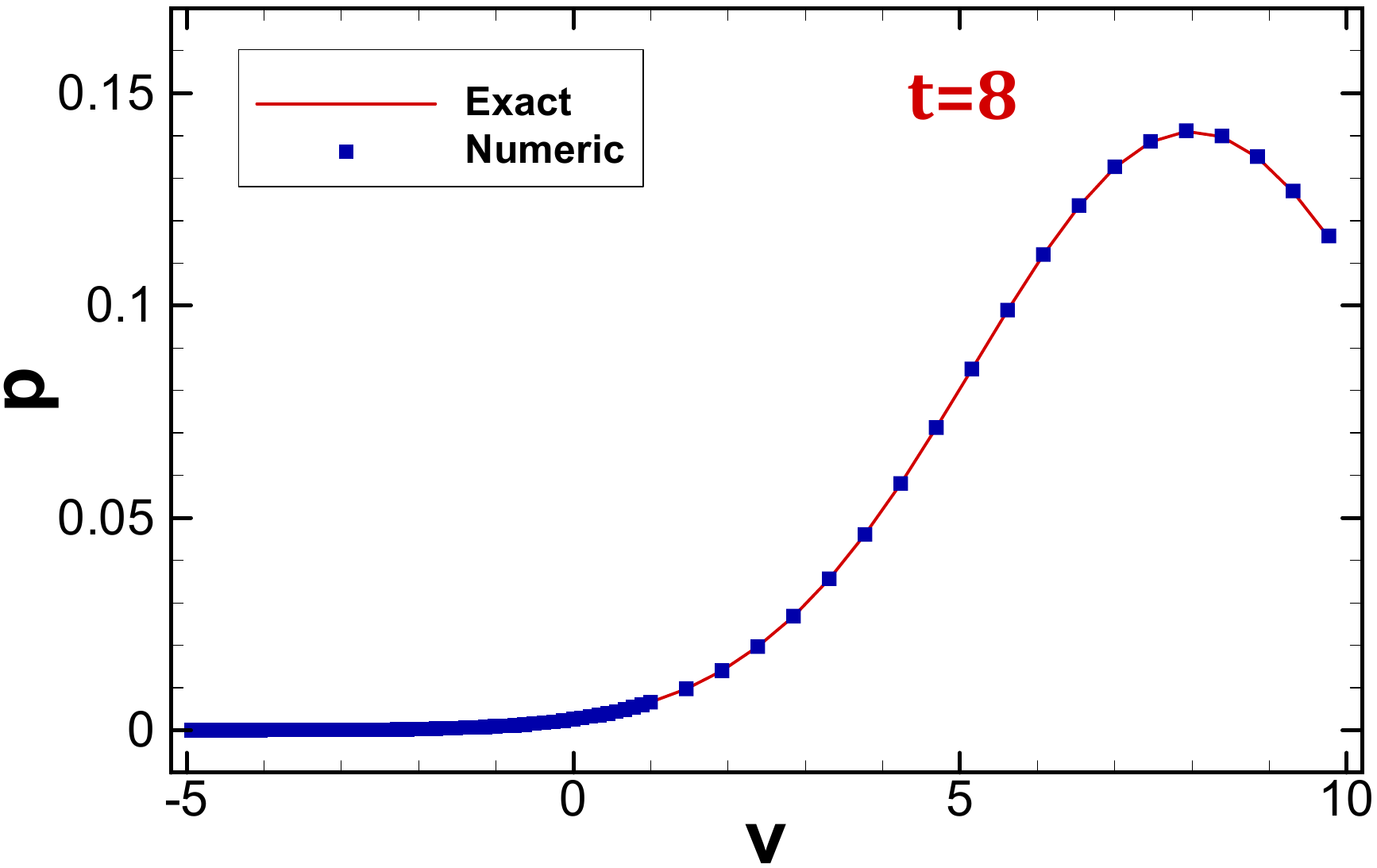}
   \end{center}
   \caption{Numerical result of Eq.~(\ref{ac}) with $\Phi(v)=1$ at time $t=8$. Here, $N_1=40$, $N_2=10$ and $N_3=20$ \blue{ are used to compute the numerical result, which matches well with the exact solution (\ref{eq_constant}). } }
      \label{fig_constant}
\end{figure}

When $\Phi(v)=\mu$ with $\mu$ being a constant, Eq.~(\ref{aa}) corresponds to the Brownian motion with a constant drift, whose propagator is simply Gaussian,
\begin{equation}
   p(v,t|v_0,0)=\frac{1}{ \sqrt{4\pi Dt}  }e^{ -(v-v_0-\mu t)^2/(4Dt) }.
   \label{eq_constant}
\end{equation}
Here $\mu=1$ and $v_0=0$ are chosen and the solution domain is truncated to be
$[-5,10]$. While a zero current boundary condition is set for the left boundary, i.e., $f(-5,t|v_0,0)=0$, no boundary condition is needed for the right due to the positiveness of the chosen $\mu$. The numerical results obtained in Tab.~\ref{tab_constant} show that for this smooth case the algorithm proposed in this paper achieves a fifth-order convergence rate approximately, \blue{ while only approximately second-order for the Chang-Cooper scheme \cite{ChangCooper1970} (see Appendix \ref{app_sec1}), which is one of the most popular schemes for solving Fokker-Planck equations.} As we can see from Fig.~\ref{fig_constant}, when $t=8$ the current is much large than zero at the right boundary\red{. But} the proposed scheme still produces a solution that matches with the exact solution.
This means that the computational domain is not necessary to be large to avoid boundary refection here. It is no doubt that this property is very desirable in numerical simulations.

\begin{table}
   \begin{tabular*}{\linewidth}{@{\extracolsep{\fill}}cccccccc}
      \hline
       \multicolumn{8}{c}{The current method} \\
      \hline
      $N_1$ &$N_2$ &$N_3$ & $N_v$ & $L^2$ error & rate & $L^\infty$ error & rate \\
      \hline
      40 & 10 & 20 & 68 & 1.39E-03 & -- &	1.94E-03\\
      80 & 20 & 40 & 138	& 5.16E-05 &4.66& 	5.21E-05&	5.11\\
      160 & 40 & 80 & 278&	1.77E-06&	4.82 &	1.46E-06	&5.11\\
      320 & 80 & 160 & 558	 & 5.09E-08	& 5.09 &4.04E-08	& 5.15
\\\hline
       \multicolumn{8}{c}{ \blue{The Chang-Cooper scheme} } \\
      \hline
       -- &-- &-- & \blue{ $N_v$ } & \blue{$L^2$ error} & \blue{rate} & \blue{$L^\infty$ error } &
        \blue{ rate } \\
      \hline
       --& -- & -- & 68  & 6.73E-01 & -- &	9.35E-01\\
       --& -- & -- & 138 & 2.50E-01 &1.40& 	3.25E-01	&1.49\\
       --& -- & -- & 278 & 3.80E-02& 2.69 &	5.80E-02	&2.46\\
       --& -- & -- & 558 & 1.02E-02	& 1.89 &1.58E-02	& 1.87
\\\hline
   \end{tabular*}
   \caption{Accuracy test for Eq.~(\ref{ac}) with $\Phi(v)=1$ at time $t=1$.
   The total number of solution points is $N_v=N_1+N_2+N_3-2$. \blue{The Chang-Cooper scheme for Eq.~(\ref{ac}) with constant drift $\Phi(v)=\mu$ is presented in Appendix \ref{app_sec1}.
   Here, the errors are computed according to Eqs.~(\ref{eq_l2error}) and (\ref{eq_linfty}), and the rates are defined by Eq.~(\ref{eq_rate}).}}
   \label{tab_constant}
\end{table}

\subsubsection{Ornstein-Uhlenbeck process}

When $\Phi(v)=-\gamma v$ with $\gamma$ being a constant, Eq.~(\ref{aa}) corresponds to the Ornstein-Uhlenbeck process.
In this case, it is well known that Eq.~(\ref{ac}) admits the solution
\begin{equation}
   p(v,t|v_0,0)=\sqrt{ \frac{\gamma}{2\pi D (1-e^{-2\gamma t}) } }\exp\left( -\frac{\gamma ( v-e^{ -\gamma t } v_0 )^2 }{ 2D( 1-e^{-2\gamma t}) } \right).
   \label{eq_ou}
\end{equation}
For $\gamma>0$ and $t\rightarrow \infty$, the solution tends to the stationary solution
\begin{equation}
  p_{_{\mathrm{OU}}}(v)=\sqrt{ \frac{\gamma}{2\pi D } }e^{-\gamma v^2/(2D) }.
\end{equation}
For $\gamma\leqslant 0$, no stationary solution exists.

\begin{figure}
   \begin{center}
       \includegraphics[width=0.9\linewidth]{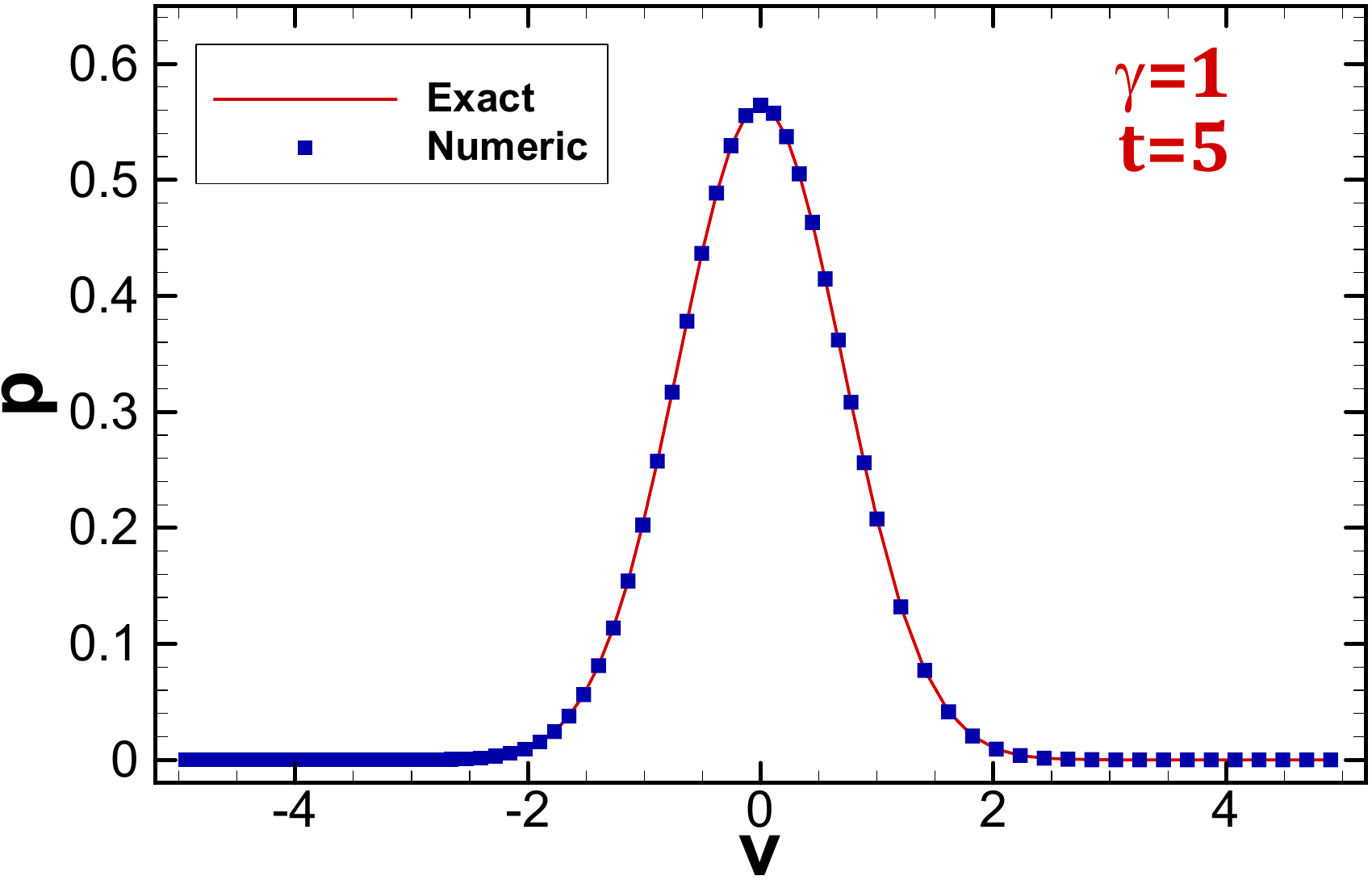}\\\vspace{0.5em}
       \includegraphics[width=0.9\linewidth]{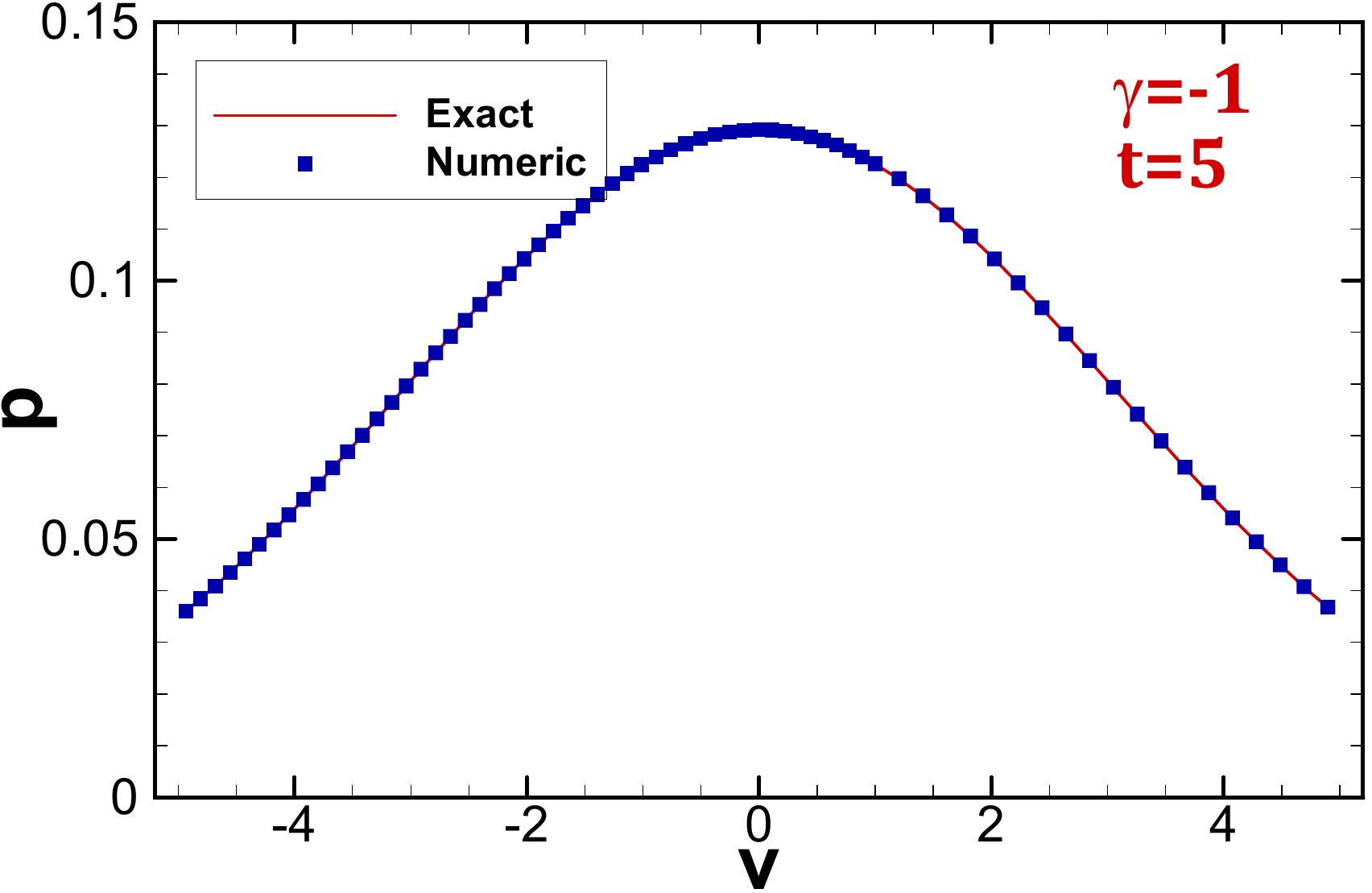}
   \end{center}
   \caption{Numerical results of Eq.~(\ref{ac}) with $\Phi(v)=-\gamma v$ for $\gamma=1$ and $\gamma=-1$ \blue{at different time}. Here, $N_1=40$, $N_2=10$ and $N_3=20$ \blue{ are used to compute the numerical results, which match well with the exact solution (\ref{eq_ou})}. }
      \label{fig_ou}
\end{figure}

Here we consider computations for the two cases $\gamma=1$ and $-1$. The computational domain $[-5,5]$ is chosen for both the cases. While zero current boundary conditions are set for the first case, no particular boundary condition is needed for the second. As we can see in Tab.~\ref{tab_ou} that the proposed scheme achieves fifth-order accuracy approximately for the two cases. Numerical results for larger time as shown in Fig.~\ref{fig_ou} confirm that the scheme is also \red{valid} for long time simulations, even for negative $\gamma$ without a large computational domain.
\begin{table}
   \begin{tabular*}{\linewidth}{@{\extracolsep{\fill}}cccccccccccccc}
      \hline
      &&&& \multicolumn{4}{c}{$\gamma=1$}\\
      \cline{5-8}
      $N_1$ &$N_2$ &$N_3$ & $N_v$ & $L^2$ error & rate & $L^\infty$ error & rate \\
      \hline
      40 & 10 & 20 & 68 & 4.11E-04	& -- &	3.93E-04 &--\\
      80 & 20 & 40 & 138	& 5.18E-05	& 2.93 &	5.63E-05	& 2.74
\\
      160 & 40 & 80 & 278&	2.06E-06	&4.60 	&2.03E-06	&4.75
\\
      320 & 80 & 160 & 558	 & 4.10E-08&	5.62 	&3.86E-08	&5.69
\\\hline
      &&&& \multicolumn{4}{c}{$\gamma=-1$}\\
      \cline{5-8}
      $N_1$ &$N_2$ &$N_3$ & $N_v$ & $L^2$ error & rate & $L^\infty$ error & rate \\
      \hline
      40 & 10 & 20 & 68 & 3.29E-04	 & -- &	2.87E-04 &--
\\
      80 & 20 & 40 & 138	& 5.59E-05	& 2.50	& 4.53E-05&	2.61
\\
      160 & 40 & 80 & 278&	2.34E-06	& 4.53	& 1.73E-06&	4.67
\\
      320 & 80 & 160 & 558	 & 4.75E-08	& 5.60	& 3.36E-08 &	5.65
\\\hline
   \end{tabular*}
   \caption{Accuracy test for Eq.~(\ref{ac}) with $\Phi(v)=-\gamma v$ at time $t=0.5$ \blue{for two different values of $\gamma$}. Here $N_v=N_1+N_2+N_3-2$. \blue{The errors are computed according to Eqs.~(\ref{eq_l2error}) and (\ref{eq_linfty}), and the rates are defined by Eq.~(\ref{eq_rate}).} }
   \label{tab_ou}
\end{table}

\subsection{Drifts admitting one jump}

Since drifts admitting only one jump are considered, the computational domain is only divided into two subdomains by the jump \red{here}. Then we can modify the proposed finite difference scheme just by removing the subdomain $\Omega_2$ as shown in Fig.~\ref{fig_grid}. Here the time step is chosen appropriately to be $\tau=0.01 \min\{h_1^2,h_2^2\}$.

\subsubsection{Pure dry friction}\label{sec_4bc}

\begin{figure}
   \includegraphics[width=0.9\linewidth]{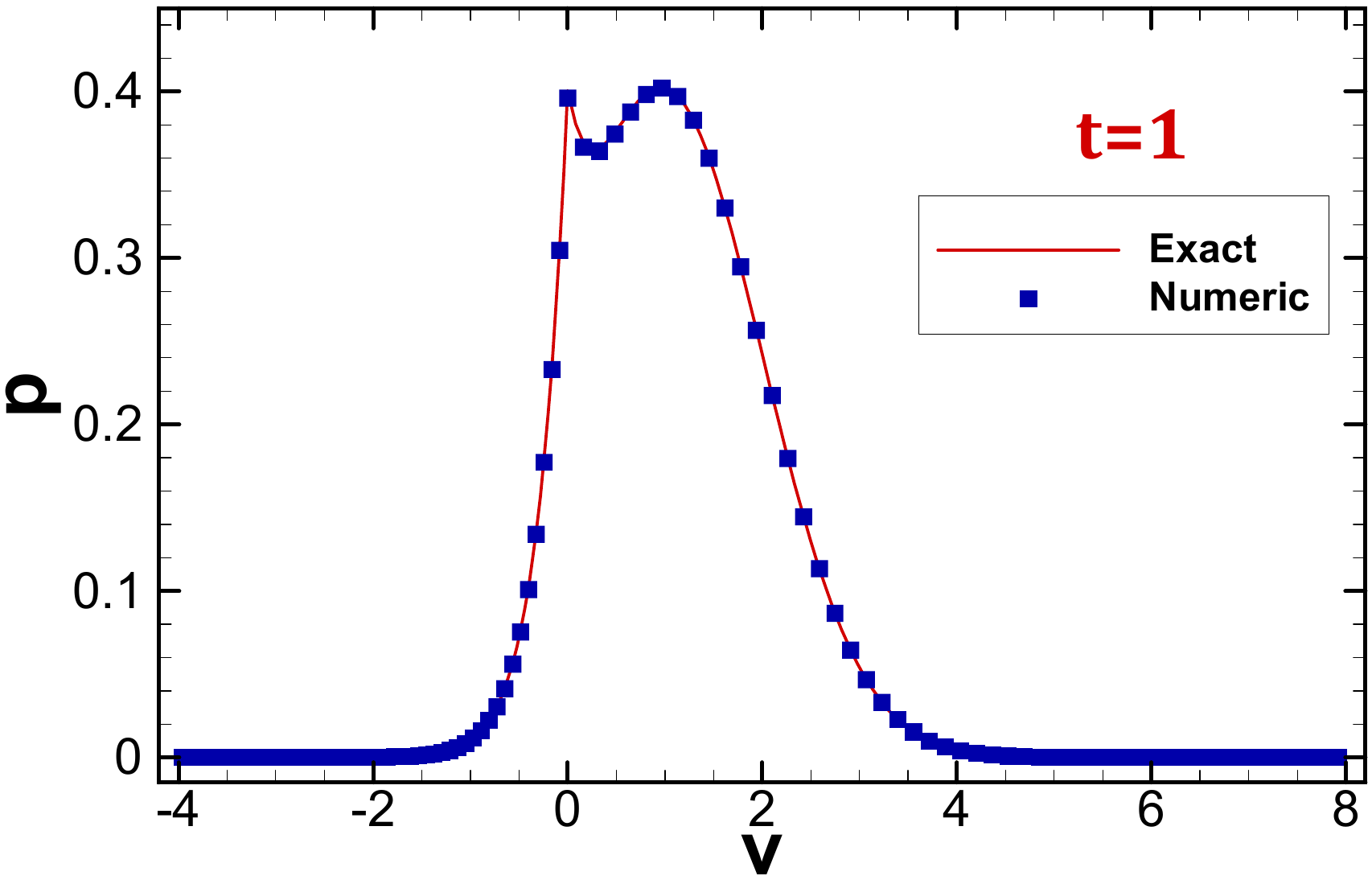}\\\vspace{0.5em}
   \includegraphics[width=0.9\linewidth]{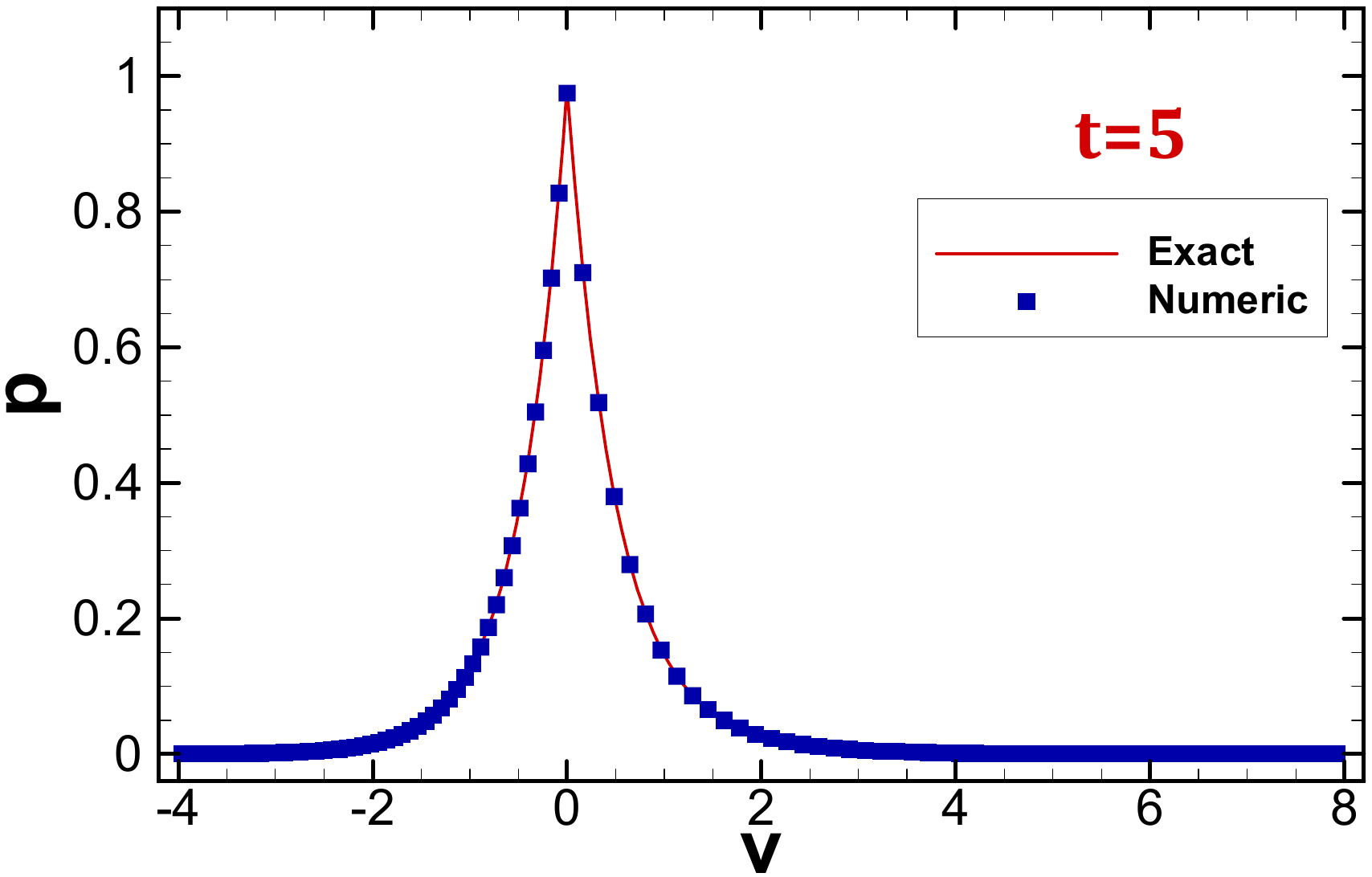}
   \caption{Propagators of \blue{Eq.~(\ref{aa}) with the drift $\Phi(v)= -\mathrm{sgn}(v)$} at $t=1$ and $t=5$. Here $v_0=2$ and $N_1=N_2=50$ \blue{are chosen to compute the numerical results, which
   matches well with the exact solution (\ref{eq_dry_sol}).} }
   \label{fig_dry}
\end{figure}

When $\Phi(v)=-\mu\, \text{sgn}(v)$ \red{with $\mu$ being positive constant and ``sgn" denoting the sign function}, Eq.~(\ref{aa}) corresponds to the Brownian motion with pure dry friction \cite{Gennes2005dryfriction}, whose propagator is known in closed analytic form \cite{CaugheyDienes1961,Karatzas1984,TouchetteStraeten2010Brownian}
\begin{align}
   p(v,t|v_0,0)=\tfrac{\mu}{D}\hat{p}\left( \tfrac{\mu}{D}v,\tfrac{\mu^2}{D}t\big|\tfrac{\mu}{D}v_0,0 \right),
    \label{eq_dry_sol}
\end{align}
where
\begin{align*}
  \hat{p}(x,\tau|x_0,0)=& \frac{e^{-\tau/4}}{2\sqrt{\pi \tau}}e^{-(|x|-|x_0|)/2}e^{-(x-x_0)^2/(4\tau)}\nonumber\\
  & + \frac{e^{-|x|}}{4}\left[
    1+\mathrm{erf}\left( \tfrac{\tau-(|x|+|x_0|)}{2\sqrt{\tau}} \right)
  \right]
\end{align*}
with $
   \mathrm{erf}(x)=2\int_0^x \exp(-z^2)dz/\sqrt{\pi}
     $ denoting the error function. The exact solution at time $t=\tau_0$ is set to be the initial condition with $v_0=2$, the computational domain is chosen to be $[-4,8]$ and zero current conditions are set at the boundaries.  Figure \ref{fig_dry} shows that the numerical results are consistent to the exact solutions. Since the solution admits a cusp at $v=0$ like peakons \cite{KalischRaynaud2006},  second-order accuracy is observed in Tab.~\ref{tab_dry}, as expected. \red{In addition, the results for a more general case with a two-valued drift are presented in Appendix \ref{app_sec2}}.

\begin{table}
   \begin{tabular*}{\linewidth}{@{\extracolsep{\fill}}ccccccc}
      \hline
      $N_1$ &$N_2$ & $N_v$ & $L^2$ error & rate & $L^\infty$ error & rate \\
      \hline
      50 & 50  & 99 & 4.73E-03	&--&	2.80E-03&--
\\
      100 & 100  & 199 & 1.21E-03	&1.96 	&6.94E-04	&2.00
\\
      200 & 200 & 399 & 3.01E-04&	2.00& 	1.72E-04&	2.00
\\
      400 & 400 & 799 & 7.57E-05	&1.99 	&4.32E-05&	1.99
\\\hline
   \end{tabular*}
   \caption{Accuracy test for \blue{Eq.~(\ref{aa}) with the drift $\Phi(v)=-\mathrm{sgn}(v)$} at $t=1$. Here, $v_0=2$ and $N_v=N_1+N_2-1$. \blue{The errors are computed according to Eqs.~(\ref{eq_l2error}) and (\ref{eq_linfty}), and the rates are defined by Eq.~(\ref{eq_rate}).}}
   \label{tab_dry}
\end{table}

\subsubsection{Other drifts admitting one jump}

Additional to the pure dry friction case, we consider here
another two drifts admitting one jump, which are studied by using exact simulations
 of Eq.~(\ref{aa}) in \cite{Etore2014}
and \cite{papaspiliopoulos2016}, respectively.

\begin{figure}
   \includegraphics[width=0.9\linewidth]{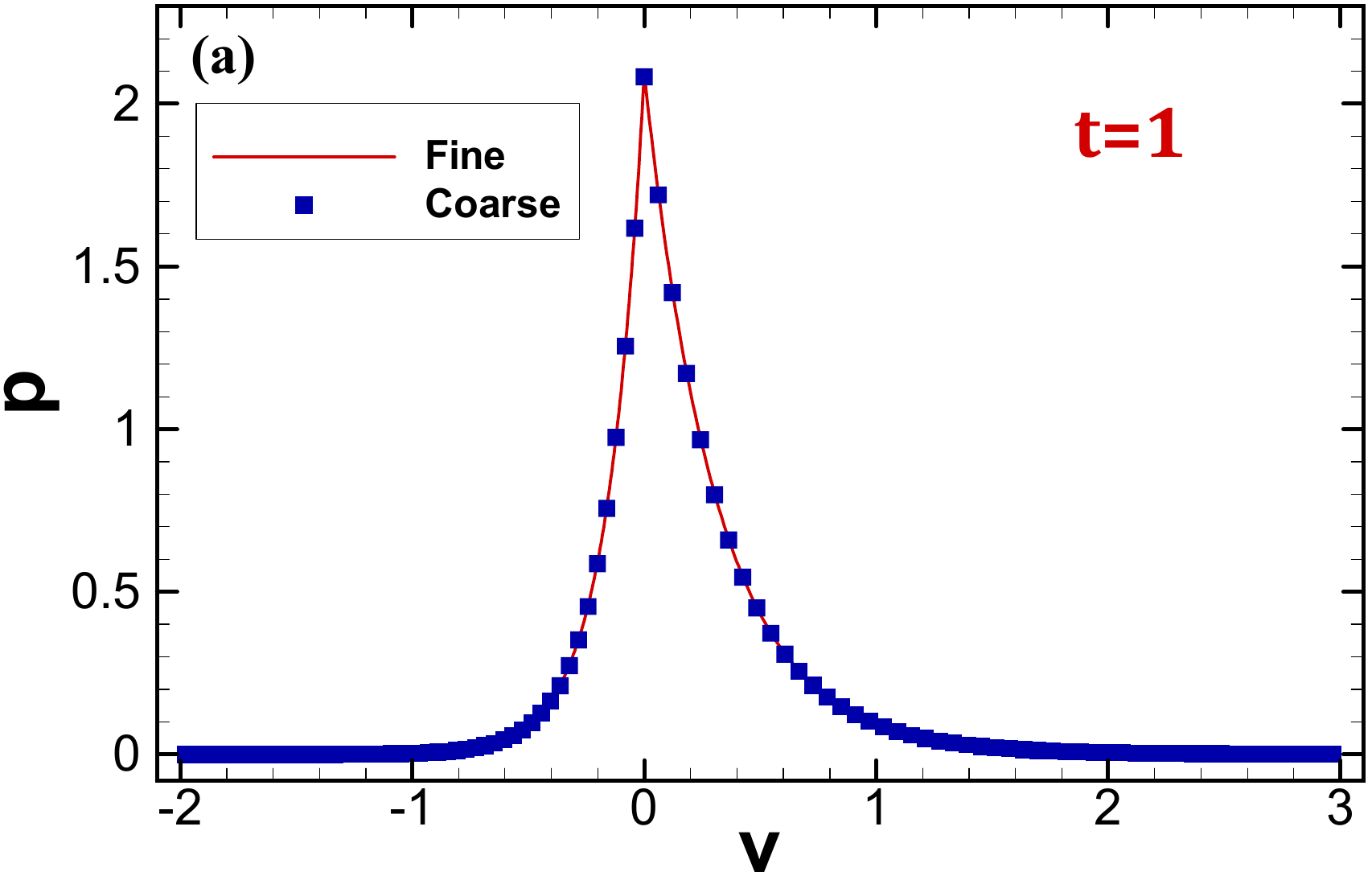}\vspace{0.5em}
   \includegraphics[width=0.9\linewidth]{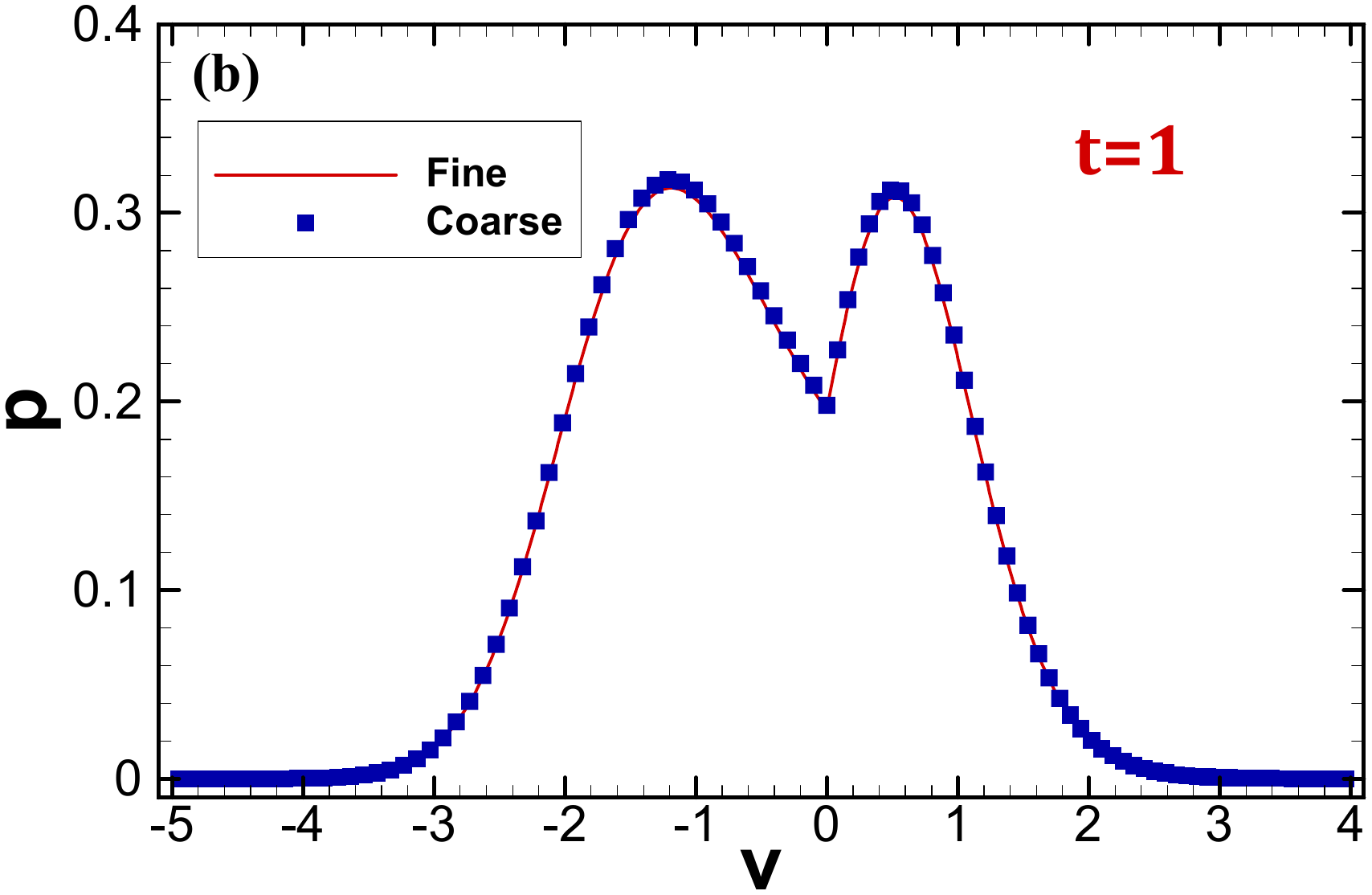}
   \caption{Propagators of Eq.~(\ref{aa}) with the drifts (\ref{eq_drift_one_r1}) \blue{ [(a)]
    and (\ref{eq_drift_one_r2}) [(b)]} at $t=1$; $v_0=0$. Here the coarse-grid solutions obtained with $N_1=N_2=50$ match well with the fine-grid solutions obtained with $N_1=N_2=400$.}
   \label{fig_one_jump_r1}
\end{figure}

The drift \red{studied in \cite{Etore2014}} is
\begin{equation}
   \Phi(v)=
   \begin{cases}
\frac{3\pi}{2}-\frac{\pi}{2}\cos\left(\frac{\pi}{5}v\right), & v<0, \\
-\frac{\pi}{2}\cos\left(\frac{\pi}{5}v\right), & v>0.
   \end{cases}
   \label{eq_drift_one_r1}
\end{equation}
We choose the computational domain to be $[-2,3]$ and impose
zero current conditions at the domain boundaries. The initial condition is set to be Eq.~(\ref{eq_v}) with $v_0=0$. However, \blue{in Eq.~(\ref{eq_drift_one_r1}) we did not define the value of $\Phi(0)$ since the proposed finite difference scheme does not involve the values of the drift at discontinuous points. Therefore, we have to define the value of $\Phi(0)$ involved in the initial condition (\ref{eq_v}). In this case, we define $\Phi(0)=[\Phi(0-)+\Phi(0+)]/2$. The same definition will be used throughout this paper.} The profile of the numerical propagator
as shown in Fig.~\ref{fig_one_jump_r1}(a) agrees with the result obtained in \cite{Etore2014} (see Fig.~4 therein). Moreover, the result obtained for a coarse grid matches with the fine-grid solution.

The drift \red{investigated in \cite{papaspiliopoulos2016}} is
\begin{equation}
   \Phi(v)=
   \begin{cases}
    \sin\left(v-\tfrac{\pi}{4}\right), & v<0, \\
     \sin\left(v-\tfrac{7\pi}{6}\right), & v>0.
   \end{cases}
   \label{eq_drift_one_r2}
\end{equation}
The solution domain is truncated to be $[-5,4]$ and zero current conditions are imposed at the domain boundaries. As we can see from Fig.~\ref{fig_one_jump_r1}(b), the results agree with that obtained in \cite{papaspiliopoulos2016} (see Fig.~1 therein).
Moreover, the results obtained by using a coarse grid and a fine grid match with each other.

\subsection{Drifts admitting two jumps}
Here, the two examples presented in \cite{Dereudre2017} are considered. In both cases, the time step is chosen to be $\tau=0.01\min\{h_1^2,h_2^2,h_3^2\}$.

\begin{figure}
   \includegraphics[width=0.9\linewidth]{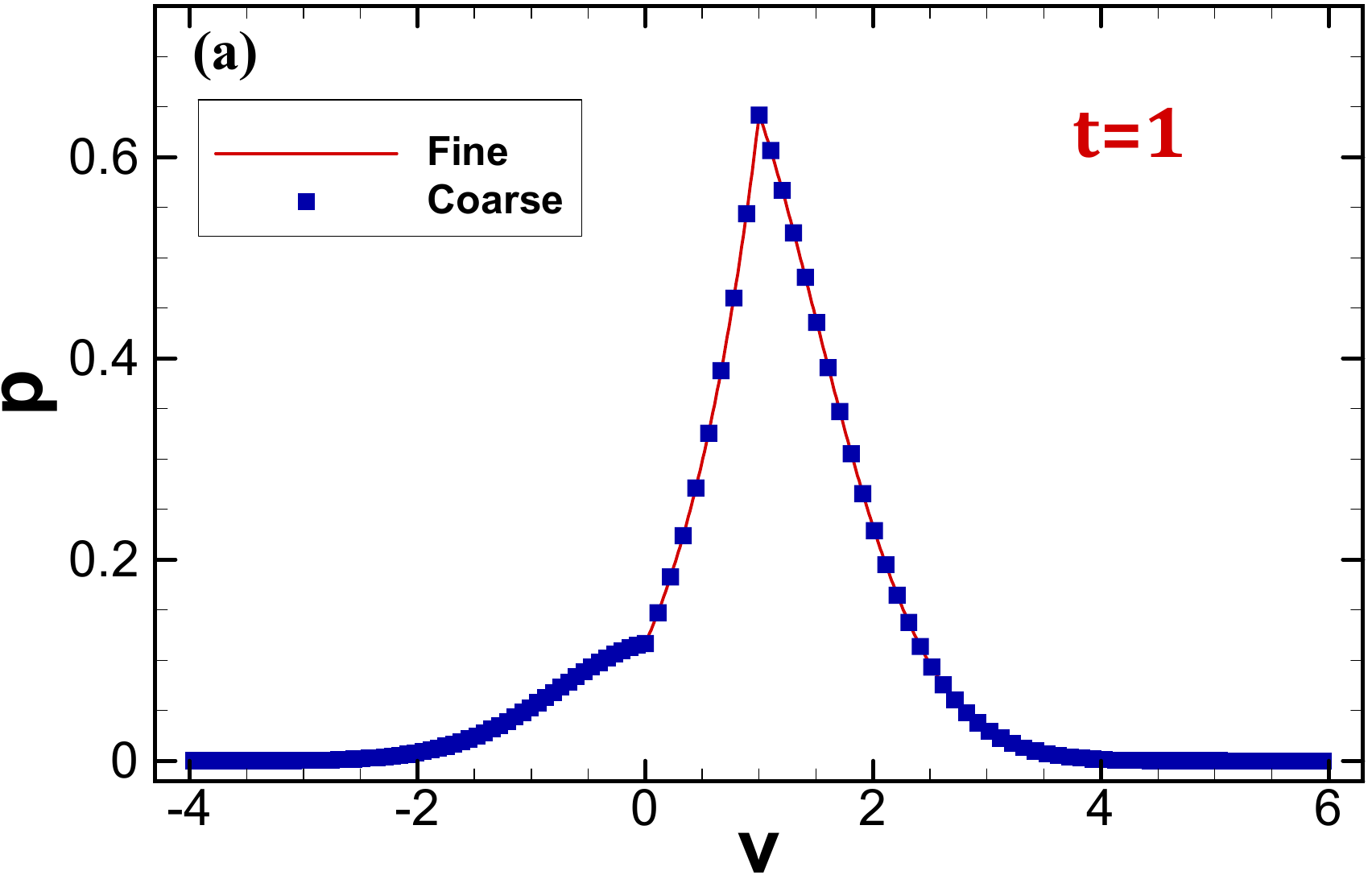}\\\vspace{0.5em}
   \includegraphics[width=0.9\linewidth]{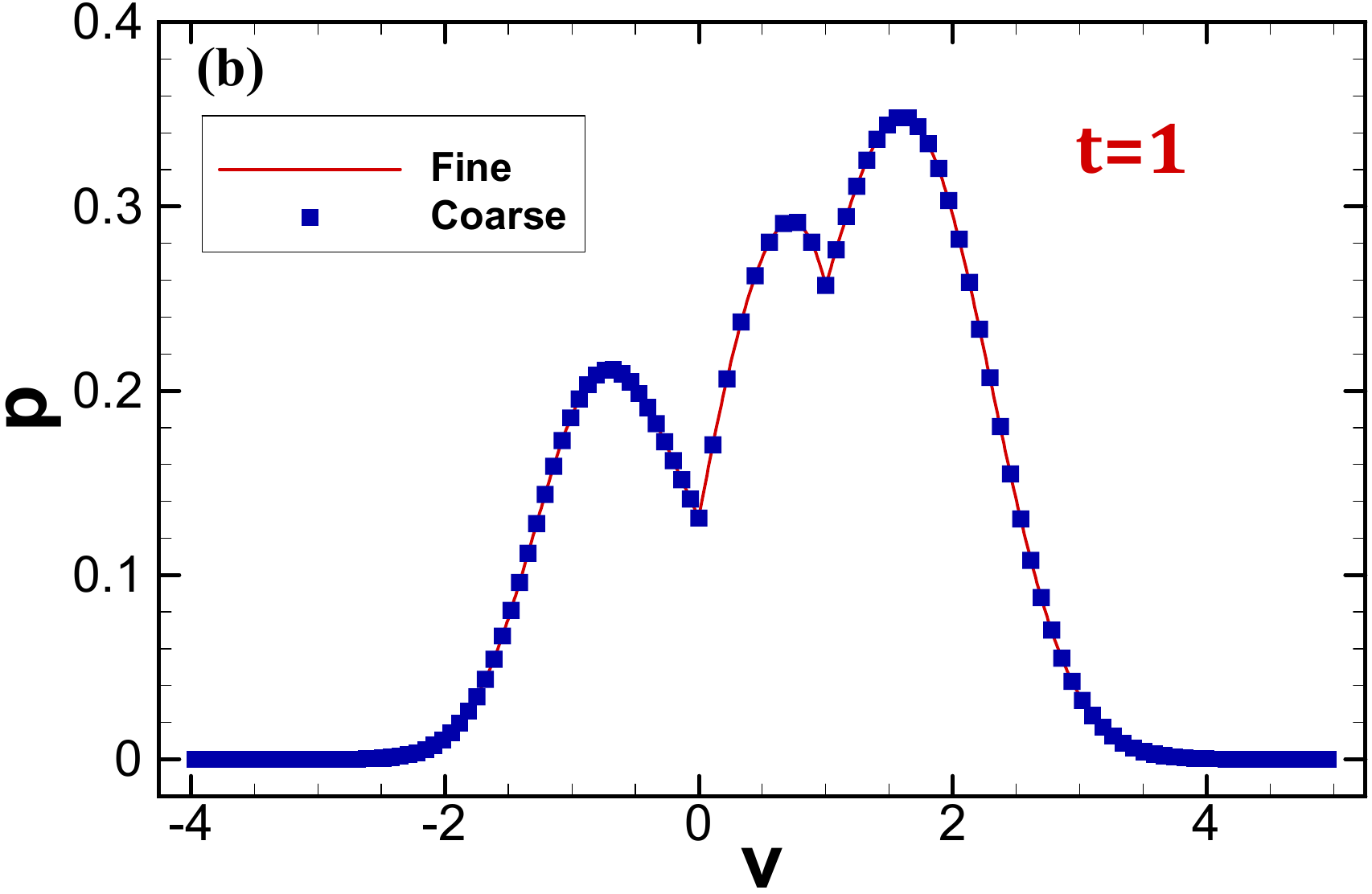}
   \caption{Propagators of Eq.~(\ref{aa}) with the drift (\ref{eq_drift_two_1}) at time $t=1$ \blue{[(a)] and the drift (\ref{eq_drift_two_2}) at time $t=0.6$ [(b)]}. Here, $v_0=0.5$ is chosen for both cases. The coarse-grid solutions are obtained with $N_1=60$, $N_2=10$ and $N_3=50$, and the fine-grid solutions are obtained with $N_1=480$, $N_2=80$ and $N_3=400$. \blue{The coarse-grid solutions match well with the fine-grid solutions.} }
   \label{fig_two_jumps_1}
\end{figure}

The first drift admitting two jumps reads as follows,
\begin{align}
   \Phi(v)=\begin{cases}
     0, & v<0,\\
     1, & 0<v<1,\\
     0, & v>1,
   \end{cases} \label{eq_drift_two_1}
\end{align}
which is piecewise-constant. In this case, the computational domain is chosen to be $[-4,6]$ and no numerical boundary condition is needed for the proposed scheme. The numerical solutions as shown
in Fig.~\ref{fig_two_jumps_1}\blue{(a)} agree with the result presented in \cite{Dereudre2017} (see Fig.~6(a) therein). In addition, the results obtained by using a coarse grid and a fine grid are consistent.

The second drift is
\begin{align}
   \Phi(v)=\begin{cases}
     -2\cos(v), & v<0,\\
     \sin(v), & 0<v<1,\\
     \cos(v-1)+\sin(1), & v>1.
   \end{cases}\label{eq_drift_two_2}
\end{align}
The computational domain is chosen to be $[-4,5]$, which is large enough for us to impose
zero current conditions at the domain boundaries for time $t=0.6$ with $v_0=0.5$. As shown in Fig.~\blue{\ref{fig_two_jumps_1}(b)}, the profile of the propagator agrees with that presented in \cite{Dereudre2017} (see Fig.~6(b) therein). Again, it is observed that the results obtained by using a coarse grid and a fine grid are consistent.

\section{Extension to functionals}
\label{sec_5}

Functionals of a stochastic process have been investigated intensively in the past and have found numerous applications in physics. Here we consider
the functional
\begin{equation}
   u(t)=\int_0^t K(v(s))ds
\end{equation}
with an integrable kernel $K(v)$, where the stochastic process $v(t)$ obeys
the Langevin equation (\ref{aa}). In particular, we have $u(0)=0$.
The joint propagator of $u$ and $v$, denoted by $p=p(u,v,t|v_0,0)$, is governed by the following Fokker-Planck equation
\begin{align}
   \partial_t\,p
   =-K(v)\partial_u\, p
   -\partial_v[ \Phi(v) p ]
   +\partial_v^2\, p
   \label{eq_two}
\end{align}
with the initial condition
\begin{equation}
   p(u,v,0|v_0,0)=\delta(u)\delta(v-v_0).
\end{equation}

\subsection{Scheme}

\begin{figure}
   \begin{center}
      \includegraphics[width=0.98\linewidth]{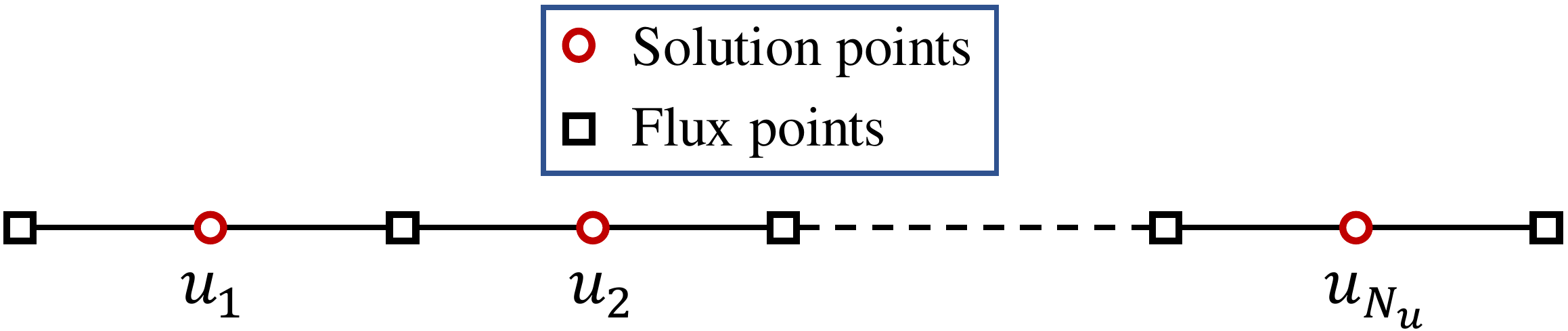}
   \end{center}
   \caption{ Illustration of the uniform grid \blue{staggered by flux points (\ref{eq_flux_u}) and solution points (\ref{eq_sol_u})} for the $u$ direction.  }\label{fig_grid_x}
\end{figure}

To solve Eq.~(\ref{eq_two}), we have to use the same matching conditions at the discontinuities of the drift $\Phi(v)$ (in the $v$ direction) as Eq.~(\ref{ac}), while in the $u$ direction we just need to use the continuous condition as usual.
Therefore, the scheme derived for Eq.~(\ref{ac}) can be applied directly for the $v$ direction.
In the $u$ direction, we choose the computational domain to be $[u_{_L},u_{_R}]$. Then a uniform staggered grid as shown in Fig.~\ref{fig_grid_x} is used. Here the flux points and solution points are
\begin{align}
   &u_{k+1/2}= u_{_L}+k h_u, \quad 0\leqslant k \leqslant N_u, \label{eq_flux_u}\\
   &u_{k}=u_{_L}+(k-1/2)h_u, \quad 1\leqslant k \leqslant N_u, \label{eq_sol_u}
\end{align}
where $N_u$ denotes the number of solution points in the $u$ direction and
the step $h_u=(u_{_R}-u_{_L})/N_u$.

Similar to
the approximation to the term $\Phi(v)p$ appearing in the current (\ref{ad}), we
approximate the term $K(v)\partial_u p$ at the point $(u_k,v_j)$ by
\[
   \min\{ K(v_j),0 \}(\partial_u p)_{k,j}^+ +\max\{ K(v_j),0 \}(\partial_u p)_{k,j}^-,
\]
 where $(\partial_u p)_{k,j}^+$ and $(\partial_u p)_{k,j}^-$ are obtained by the following procedure. For a fixed $v_j$, we first reconstruct the values of the propagator at $u_{k+1/2}$ from
the values at solution points by using fifth-order interpolations. Introducing the following two $(N_u+1)\times N_u$ interpolation matrices,
\begin{equation*}
   I_u^-=
   \begin{bmatrix}
   (\mathbf{a})_{1\times N_u}  \\[3pt]
   (I_3^-)_{N_u\times N_u}
   \end{bmatrix},\quad
      I_u^+=
   \begin{bmatrix}
   (I_1^+)_{N_u\times N_u} \\[3pt]
    \widetilde{\mathbf{a}}
   \end{bmatrix},
\end{equation*}
where $\mathbf{a}$ is defined in Eq.~(\ref{eq_def_b}) and
$\widetilde{\mathbf{a}}$ is defined by letting its entries satisfy that
$
  [\widetilde{\mathbf{a}}]_{1,j}=[\mathbf{a}_{1\times N_u}]_{1,N_u+1-j}
$, we can express the reconstructions as
$
   \mathbf{p}_{\bullet j}^{\pm}=I_u^\pm \mathbf{p}_{\bullet j},
$
where \red{the vectors} $\mathbf{p}_{\bullet j}^\pm = [ p_{1/2,j}^\pm,p_{3/2,j}^\pm,\dots,p_{N_u+1/2,j}^\pm]^{T} $ and $\mathbf{p}_{\bullet j} = [ p_{1,j},p_{2,j},\dots,p_{N_u,j}]^{T} $.
Then we approximate the derivative $\partial_u p$ at solution points as
$
   (\partial_u p)_{\bullet j}^\pm=A_u\mathbf{p}_{\bullet j}^{\pm}/h_u,
$
where the vectors $(\partial_u p)_{\bullet j}^\pm = [ (\partial_u p)_{1,j}^\pm,(\partial_u p)_{2,j}^\pm,\dots,(\partial_u p)_{N_u,j}^\pm ]^{T} $ and the $N_u\times (N_u+1)$ difference matrix $A_u$ is defined by
$
  \blue{
   A_u=(A_2)_{(N_u-1)\times N_u}
   }
$
such that the difference scheme is sixth-order at $u_k$ with $3\leqslant k \leqslant  N_u-2$ and fourth-order at the other solution points.

Here, the third-order Runge-Kutta scheme (\ref{rk_1}) is still used to solve the resulting ordinary differential system.

\subsection{Displacement}

Particularly, in this work we focus on the displacement $u(t)=\int_0^t v(s)ds$ associated with the Brownian motion with pure dry friction, i.e.,
$\Phi(v)=-\text{sgn}(v)$ and $K(v)=v$ are chosen for Eq.~(\ref{eq_two}). Here the
the computational domain in the $v$ direction is divided into two subdomains by $v=0$.
We set the initial condition to be
\begin{align}
      p(u,v,\tau_0|v_0,0)=&\frac{1}{ 4\pi D \tau_0  }e^{ -(u-v_0 \tau_0)^2/(4D \tau_0) }\nonumber\\
      &\times e^{ -(v-v_0-\Phi(v_0)\tau_0)^2/(4D \tau_0)}
      \label{eq_uv}
\end{align}
and start the computations at $t=\tau_0$ \red{with} $\tau_0$ chosen to be $0.01$.
In addition, the time step is chosen to be $\tau=0.01\min\{h_u,h_1,h_2\}$.

As mentioned in the Introduction, for the Brownian motion with pure dry friction, analytic expressions of the first two moments of the displacement are available by solving a backward Komogorov equation \cite{ChenJust2014II} or using the method based on the Pugachev-Sveshnikov equation \cite{Berezin2018}. For instance,
when $v_0=0$ and $D=1$ we can inverse the expressions (70) and (73) in \cite{ChenJust2014II} to obtain the first two moments as
\begin{align}
   M_1(t)=&0,  \label{eq_mom1}\\
   M_2(t)=& \left[\tfrac{1}{8} t^4+\tfrac{5}{6} t^3-2 t^2 +6 t-10\right]\text{erfc}\left(\tfrac{\sqrt{t}}{2}\right)\nonumber\\
   & +\Big[64-\sqrt{\tfrac{t}{\pi}} \left(\tfrac{1}{4} t^3+\tfrac{7}{6} t^2- \tfrac{13}{3} t+10\right)\Big] e^{-t/4} \nonumber\\
   & +  10t-54,  \label{eq_mom2}
\end{align}
where $\text{erfc}(x)$ is the complementary error function.

\begin{figure*}
  \begin{center}
   \includegraphics[width=0.4\linewidth]{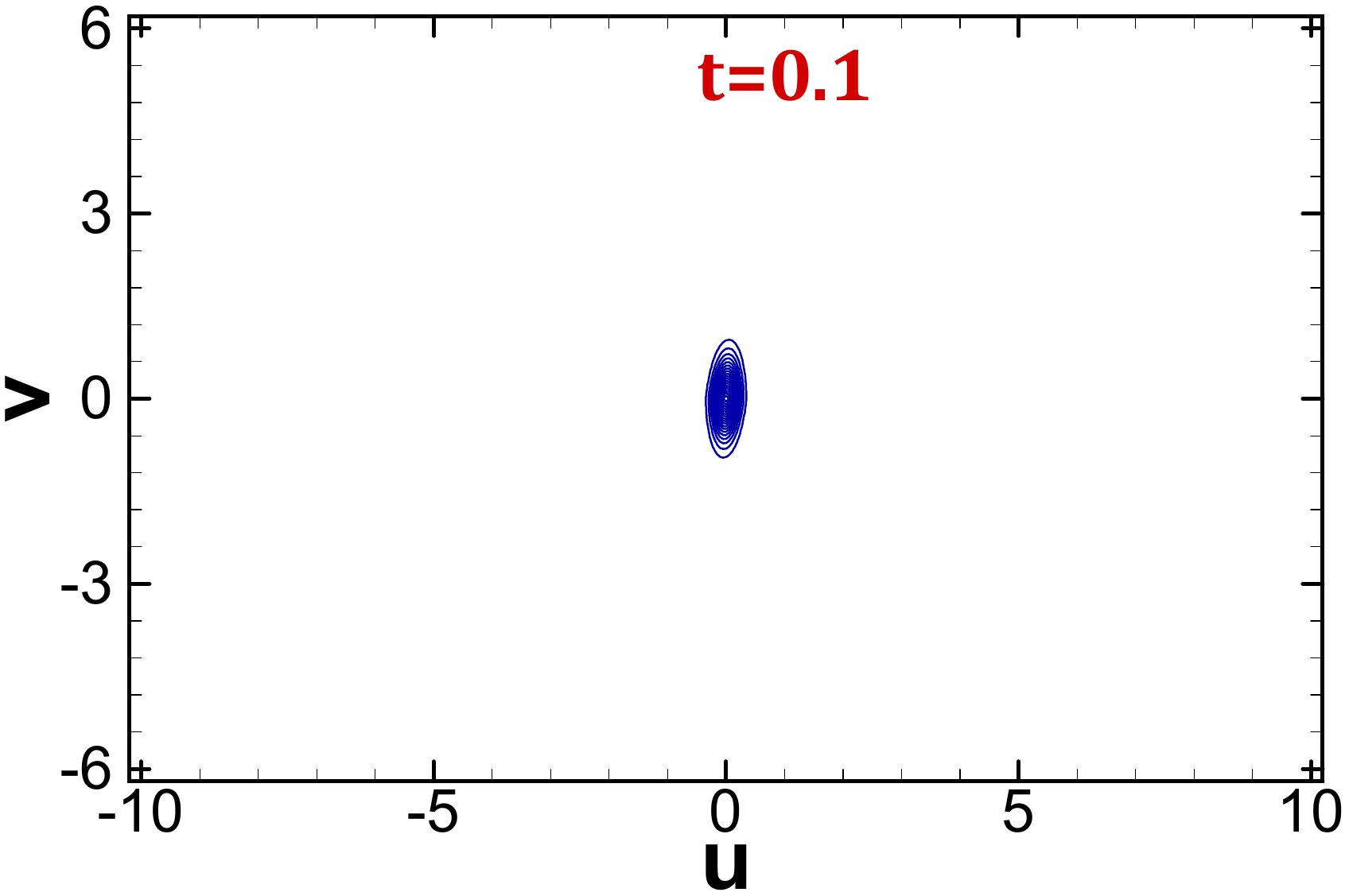}\quad
   \includegraphics[width=0.4\linewidth]{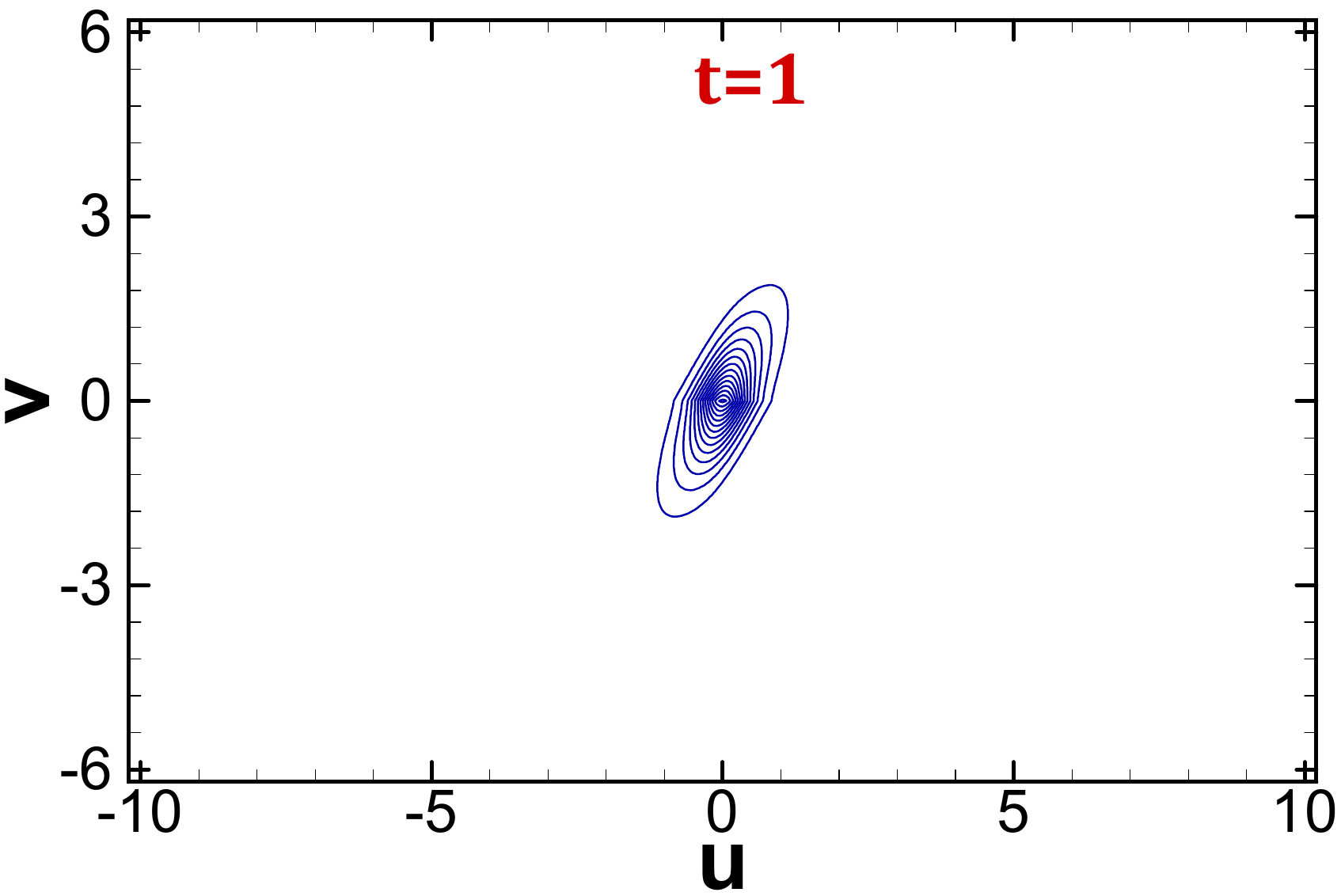} \\\vspace{0.5em}
   \includegraphics[width=0.4\linewidth]{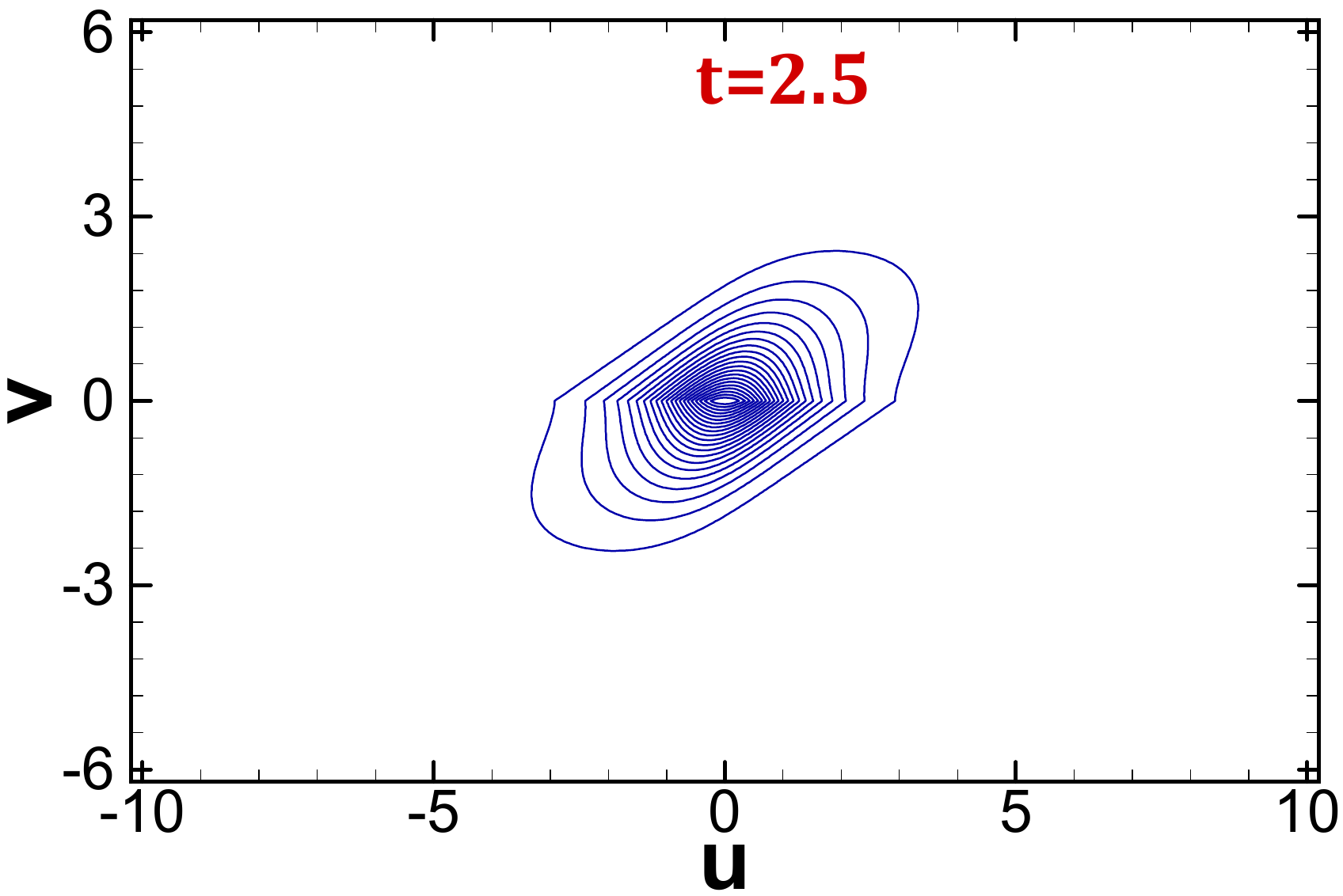}\quad
   \includegraphics[width=0.4\linewidth]{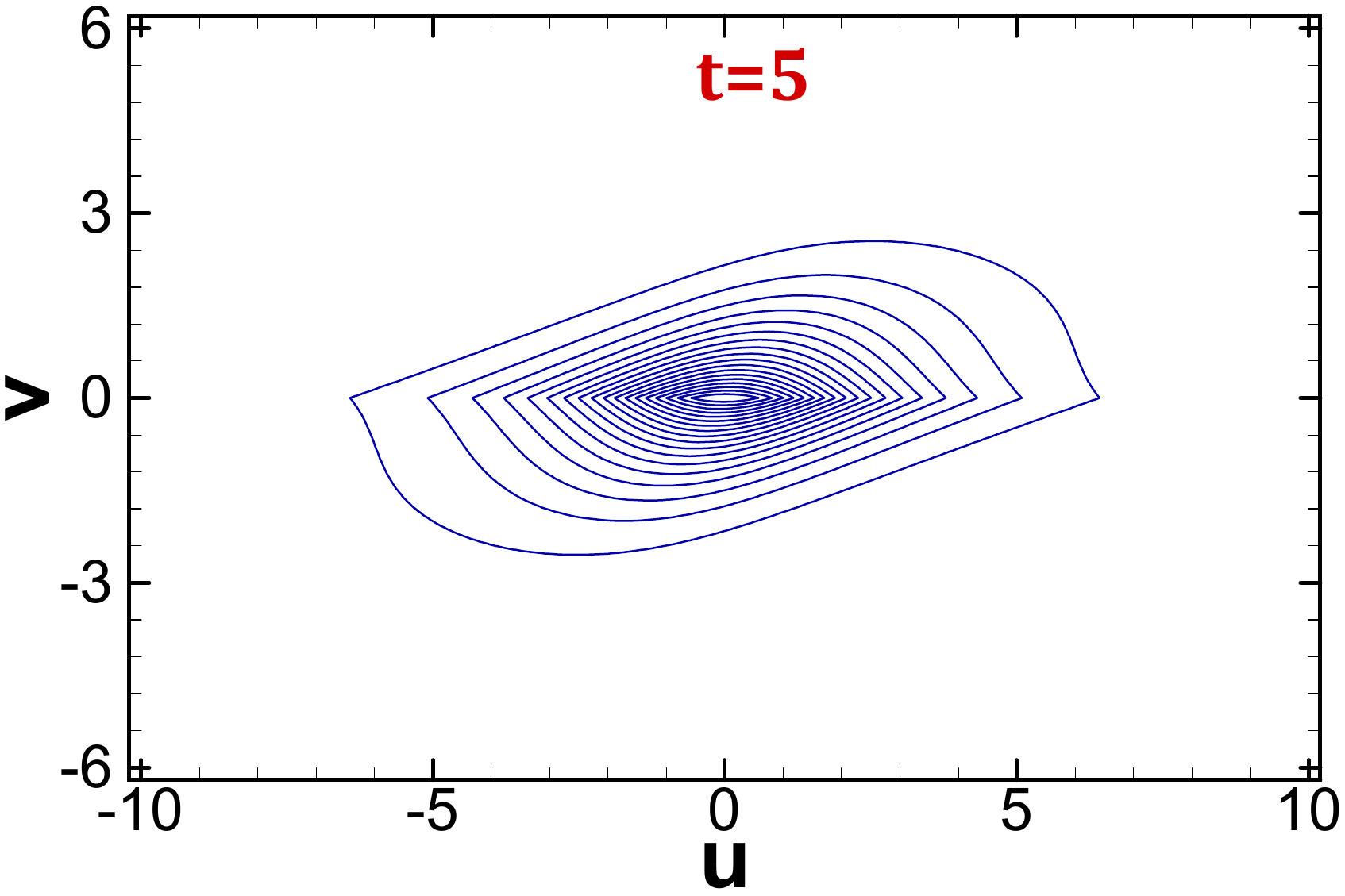}
   \end{center}
   \caption{Contour plots of the joint propagator $p(u,v,t|v_0,0)$ of Eq.~(\ref{eq_two}) at different time with $\Phi(v)=-\text{sgn}(v)$, $K(v)=v$, $D=1$ and $v_0=0$. Here $N_1=N_2=100$ and $N_u=200$ \blue{are chosen to compute the numerical results}.}
   \label{fig_diss}
\end{figure*}

\begin{figure*}
  \begin{center}
   \includegraphics[width=0.4\linewidth]{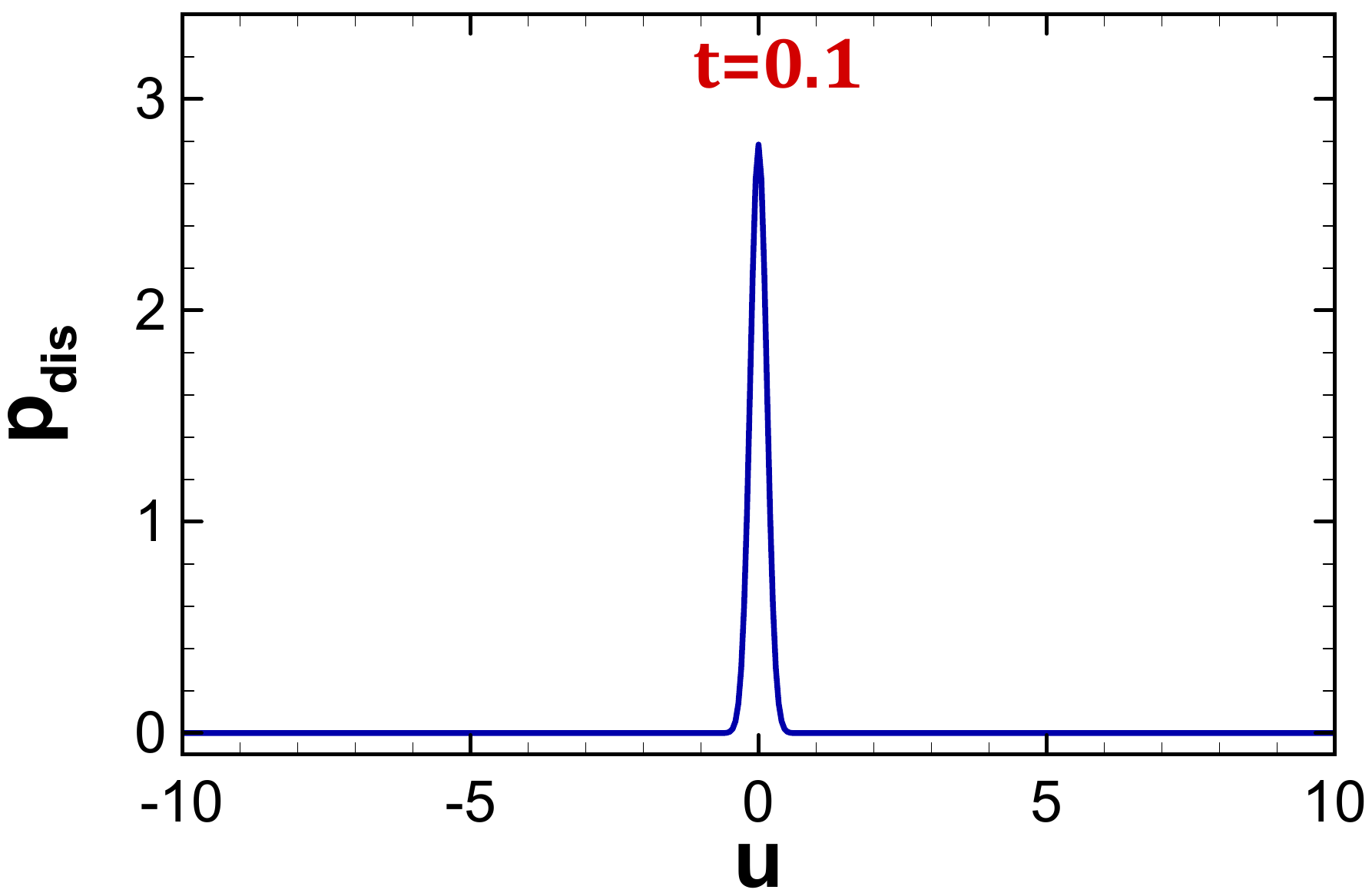}\quad
   \includegraphics[width=0.4\linewidth]{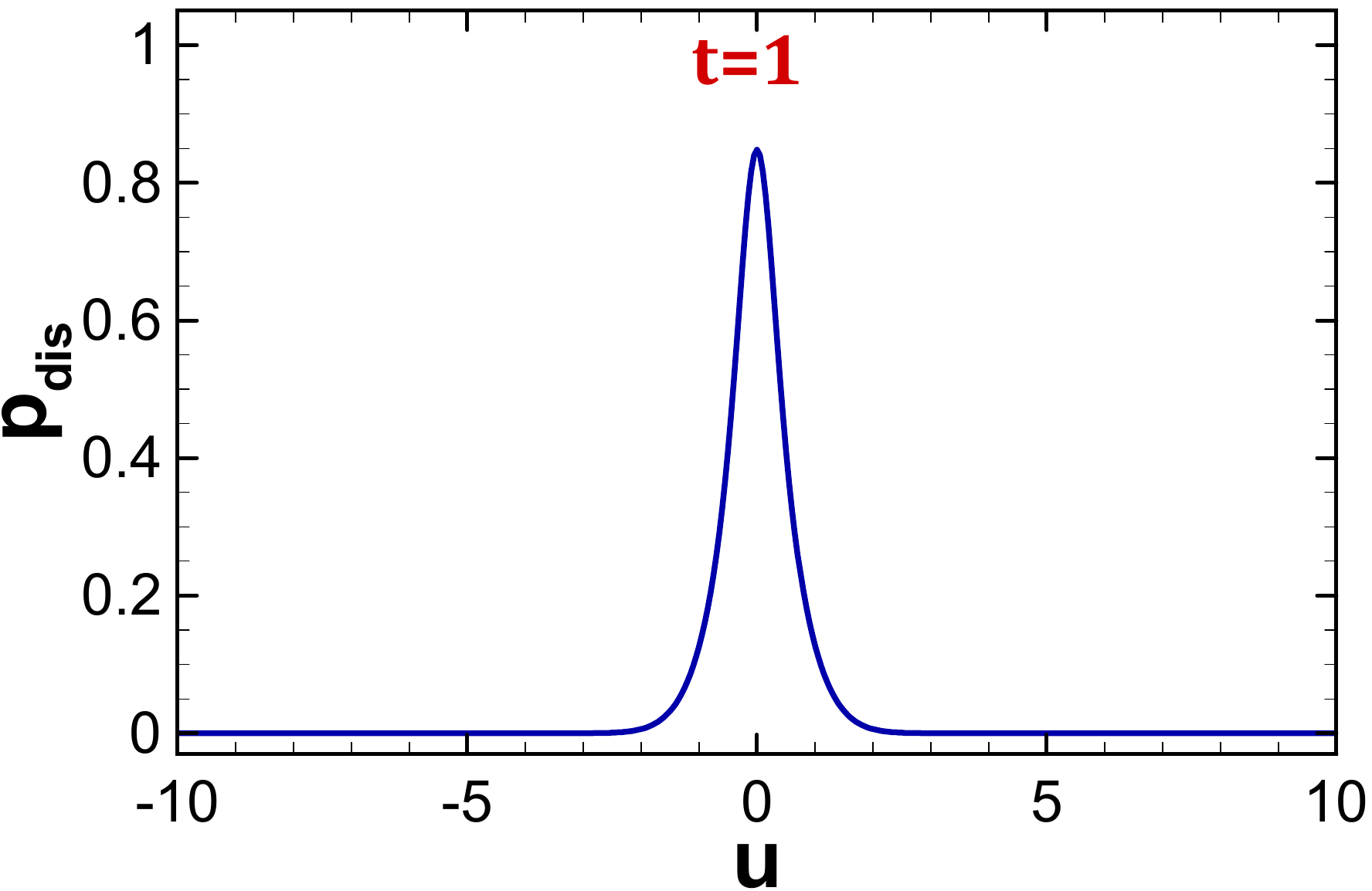}\\\vspace{0.5em}
   \includegraphics[width=0.4\linewidth]{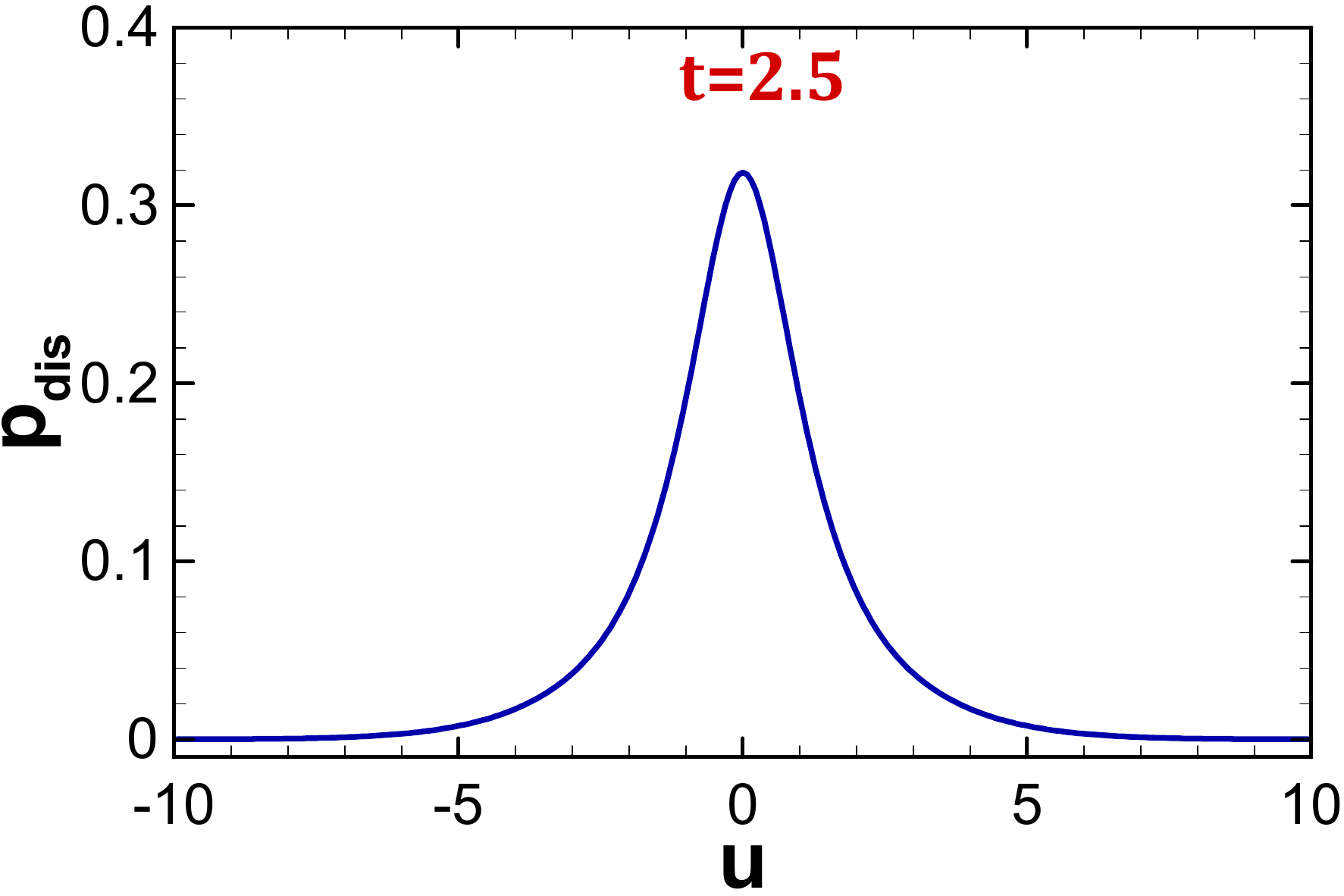}\quad
   \includegraphics[width=0.4\linewidth]{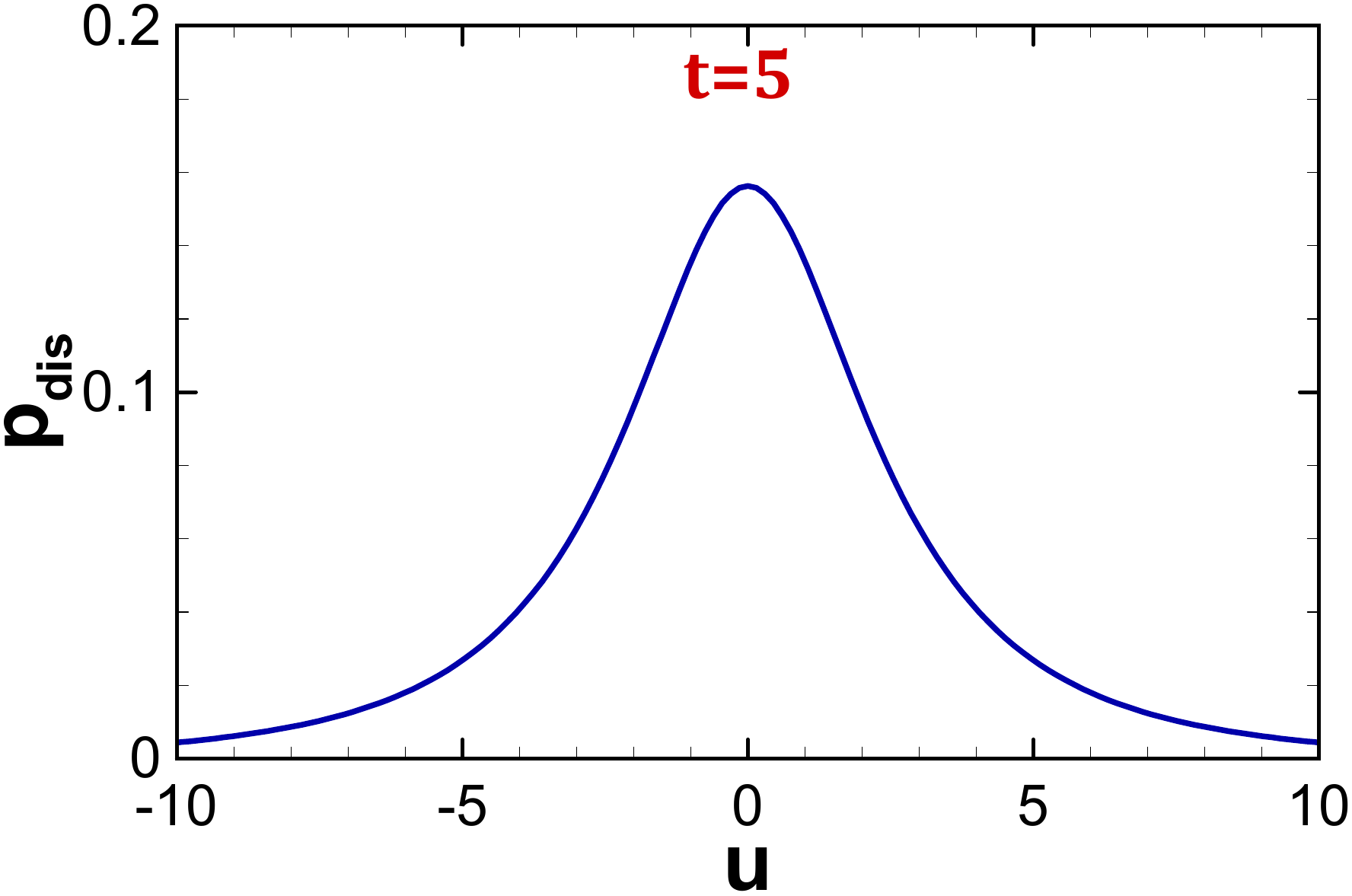}
   \end{center}
   \caption{Propagators of the displacement (\ref{eq_dis}) correspond to the numerical joint propagator
   $p(u,v,t|v_0,0)$ at different time, as shown in Fig.~\ref{fig_diss}.}
   \label{fig_dis_x}
\end{figure*}

\begin{figure*}
  \begin{center}
   \includegraphics[width=0.4\linewidth]{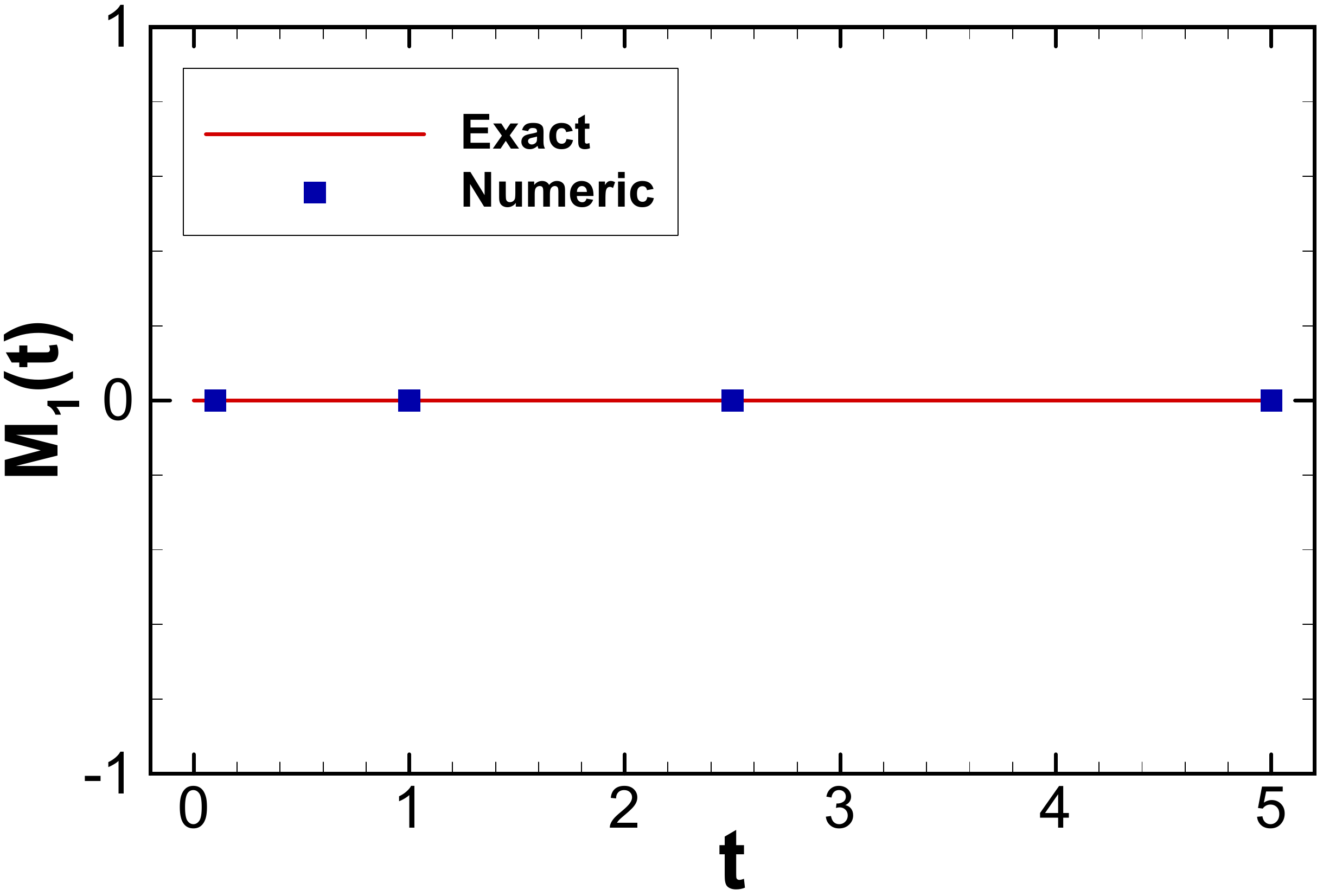}\quad
   \includegraphics[width=0.4\linewidth]{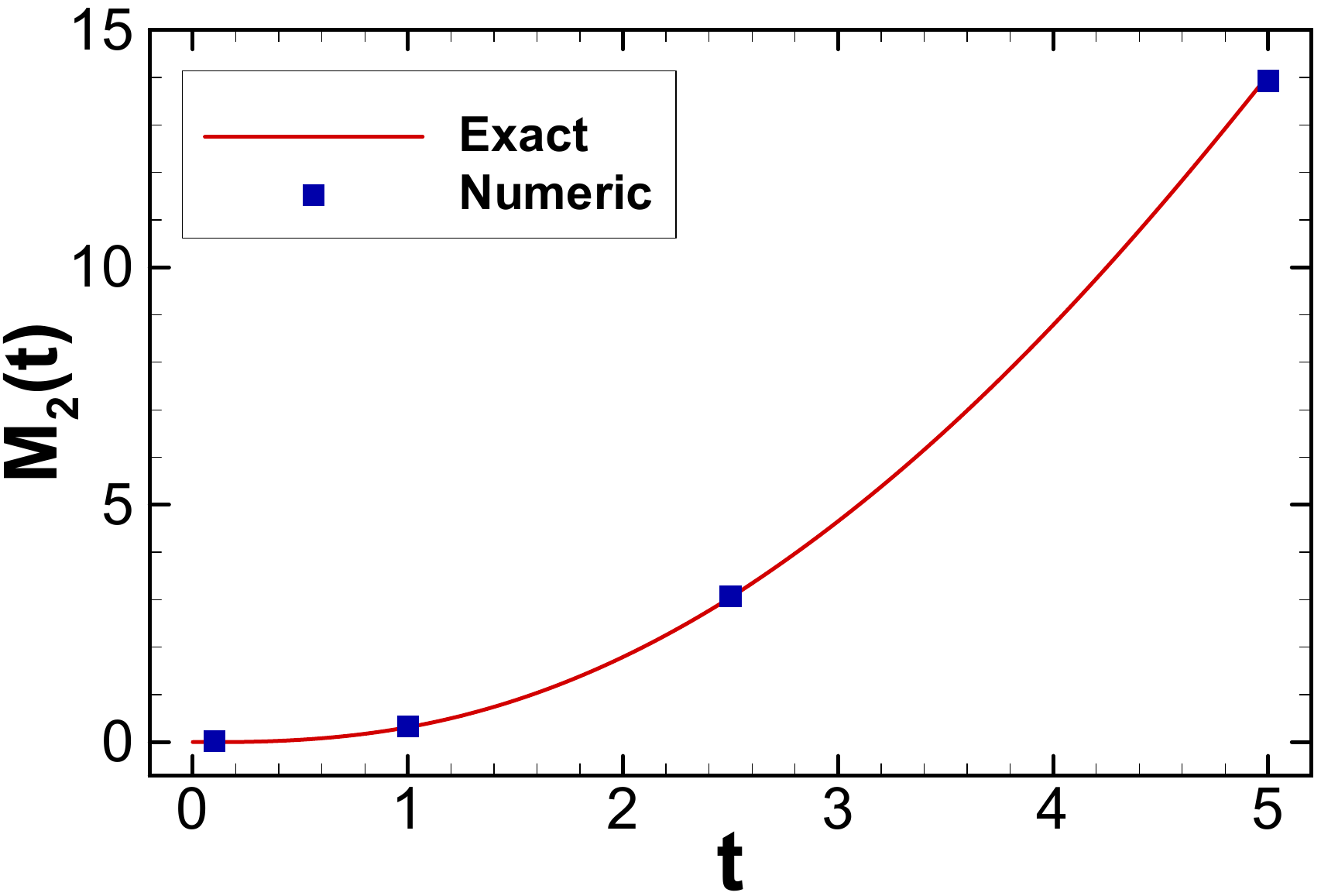}
   \end{center}
   \caption{The first two moments of the displacement. Lines correspond to the exact solutions (\ref{eq_mom1}) and (\ref{eq_mom2}), and points to the numerical solutions obtained by using Eq.~(\ref{eq_moments}) for the displacement distributions as shown in Fig.~\ref{fig_dis_x}.}
   \label{fig_diss_mom}
\end{figure*}

For different time $t$, computational domain\red{s} can be chosen differently. For simplicity, the domain in the $v$ direction is fixed to be $[-6,6]$, while the domain for the $u$ direction is set to be dependent on time. To compute the results shown in Fig.~\ref{fig_diss}, the computational domain $u\in [-10,10]$ is chosen for $t=0.1$ and $t=1$,  $u\in [-15,15]$ for $t=2.5$, and $u\in [-30,30]$ for $t=5$. While zero current conditions are set at the domain boundaries in the $v$ direction, no particular boundary condition is needed in the $u$ direction. \red{N}umerical evolution of the joined propagator is shown in Fig.~\ref{fig_diss}. In addition, the corresponding propagators of the displacement obtained by
\begin{align}
  p_{\text{dis}}( u_k,t|0 )=& \sum_{j=1}^{N_1-1}p_{j,k} h_1 +\frac{1}{2}p_{N_1,k}(h_1+h_2)\nonumber\\
  & +\sum_{j=N_1+1}^{N_1+N_2-1}p_{j,k} h_2
  \label{eq_dis}
\end{align}
are illustrated in Fig.~\ref{fig_dis_x}. To confirm the
correctness of the results, the first two moments of the displacement are computed numerically by
\begin{align}
   \widetilde{M}_s(t)=h_u\sum_{k=1}^{N_u} (u_k)^s p_{\text{dis}}( u_k,t|0 ),\quad s=1,2.
  \label{eq_moments}
\end{align}
As shown in Fig.~\ref{fig_diss_mom}, the
results agree with the analytical expressions (\ref{eq_mom1}) and (\ref{eq_mom2}), indicating the validity of the numerical method.

\section{Conclusions}
\label{sec_6}

We have derived in this paper a finite difference scheme to solve Fokker-Planck equations with drift-admitting jumps. The scheme is based on a grid staggered by flux points and solution points. In particular, the positions of the jumps are set to be solution points and used to split the solution domain into subdomains, such that we do not have to do much work to deal with the matching conditions of the propagator and the probability current at the jumps. Some benchmark problems have been computed numerically to show the validity of the scheme. The results showed that the scheme is fifth-order for the cases with smooth drifts and second-order for the cases with discontinuous drifts.

One of the desirable properties of the scheme is that, depending on the signs of the drift $\Phi(v)$ at the domain boundaries, we may not need to specify boundary conditions for the proposed scheme and could use a small computational domain to get a correct solution. This property is in particular useful when we extend the scheme to study functionals of a process, where no boundary condition is needed at the domain boundaries of the functionals. The displacement statistics of the Brownian motion with pure dry friction have been computed to show the effectiveness of the extended scheme.

The proposed numerical approach may be generalized to solve other problems involving
discontinuous drifts, e.g., problems with both discontinuous drifts and some colored noises \cite{GeffertJust2017}, and high-dimensional problems with drift-admitting jumps \cite{DasPuri2017}.

\begin{acknowledgments}
  This work was supported by the National Natural Science Foundation of China (Grant No. 11601517) and the Basic Research Foundation of National University of Defense
Technology (No. ZDYYJ-CYJ20140101).
\end{acknowledgments}

%

\appendix

\section{\blue{Chang-Cooper scheme for constant drift}}\label{app_sec1}
\blue{
Divide the computational interval $[v_{_L},v_{_R}]$ into
$N$ cells with the nodes $v_j$ satisfying
\begin{equation}
   v_j= v_{_L}+j\, h, \quad 0\leqslant j \leqslant N,
\end{equation}
where the step $h=(v_{_R}-v_{_L})/N$. Then the Chang-Cooper scheme \cite{ChangCooper1970} for the Fokker-Planck equation (\ref{ac}) with the constant drift $\Phi(v)=\mu$
can be written as
\begin{align}
   \frac{1}{\tau}(p_j^{n+1}-p_j^n)=
   &\frac{1}{h}\frac{\mu}{1-e^{-\mu h /D}}
                    (p_{j-1}^{n+1}-2p_j^{n+1}+p_{j+1}^{n+1}) \nonumber \\
                    & -\frac{\mu}{h} (p_{j+1}^{n+1}-p_j^{n+1})                    \label{aa_app}
\end{align}
for $1\leqslant j \leqslant N-1$. Here $\tau$ is the time step.
As stated in Sec.~\ref{sec_4aa} for $\mu>0$, we impose zero boundary condition $p_0^{n+1}=0$ for all $n$.
For $j=N$, we simply use an extrapolation scheme $p_N^{n+1}=2p_{N-1}^{n+1}-p_{N-2}^{n+1}$.
Now introduce the vector $\mathbf{p}^n=[p_1^n,p_2^n,\dots,p_{N-1}^n]$, we can write
the scheme (\ref{aa_app}) as a compact form
$
    Z \mathbf{p}^{n+1}=\mathbf{p}^n
$,
where the matrix
\begin{equation}
   Z= \begin{bmatrix}
       b & c \\
       a & b & c\\
       & \ddots & \ddots & \ddots \\
       & & a & b & c \\
       & & &   a-c & b+2c
    \end{bmatrix}
\end{equation}
with
\begin{align}
   a=\frac{\mu r}{ \lambda-1 },\;
   b=1-\mu r\frac{ \lambda+1 }{ \lambda-1 },\;c= \frac{ \mu r\, \lambda }{ \lambda-1 }.
\end{align}
Here $r=\tau/h$ and $\lambda=e^{-\mu h/D}$.
}

\section{Two-valued drift}\label{app_sec2}
Let us consider in Eq.~(\ref{aa}) the two-valued drift
\begin{equation}
   \Phi(v)=
   \begin{cases}
     \mu_{_L}, & v< 0,\\
     -\mu_{_R}, & v> 0,
   \end{cases}
   \label{eq_drift_two}
\end{equation}
where $\mu_{_L}$ and $\mu_{_R}$ are constants. Equation (\ref{aa}) with the drift (\ref{eq_drift_two}) is called the Brownian motion with a two-valued drift, whose propagator
can be expressed in terms of convolution integrals
 (see Eq.~(5.7) in \cite{Karatzas1984} or Eq.~(42) in \cite{SimpsonKuske2014TwoValued}).
In the following, we consider three cases according to the signs of $\mu_{_L}$ and $\mu_{_R}$. \red{Here the point $v=0$ is used to divide the computation domains into two subdomains and $\tau=0.01\min\{h_1^2,h_2^2\}$ is chosen for the time step. In addition, we set the initial condition to be Eq.~(\ref{eq_v}) with $\tau_0=0.01$ and start the computations from $t=\tau_0$.}

\subsection{Case 1: $ \mu_{_L}>0$ and $\mu_{_R}>0 $}

\begin{figure*}
   \includegraphics[width=0.45\linewidth]{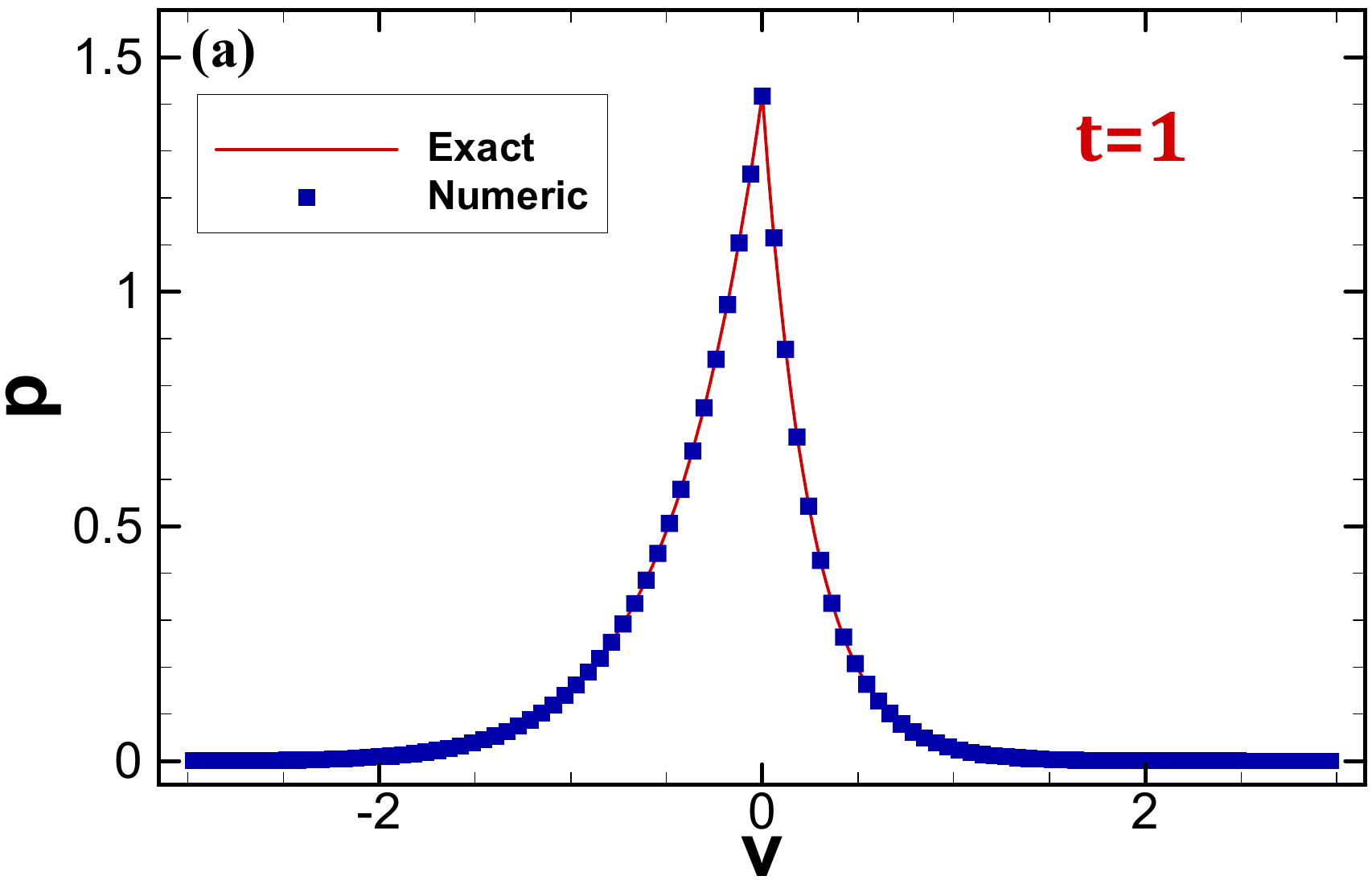}\quad
   \includegraphics[width=0.45\linewidth]{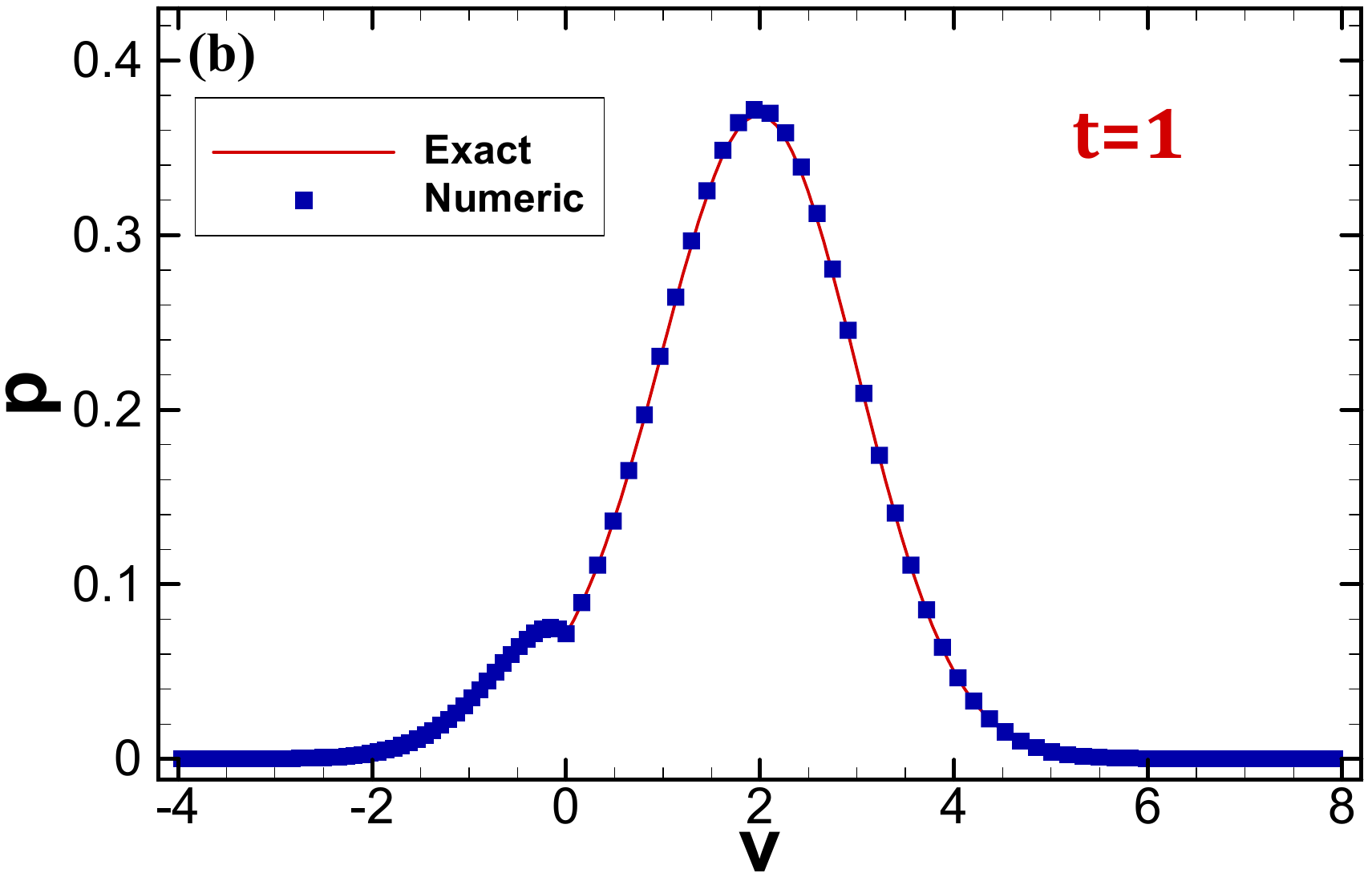}\\\vspace{0.5em}
   \includegraphics[width=0.45\linewidth]{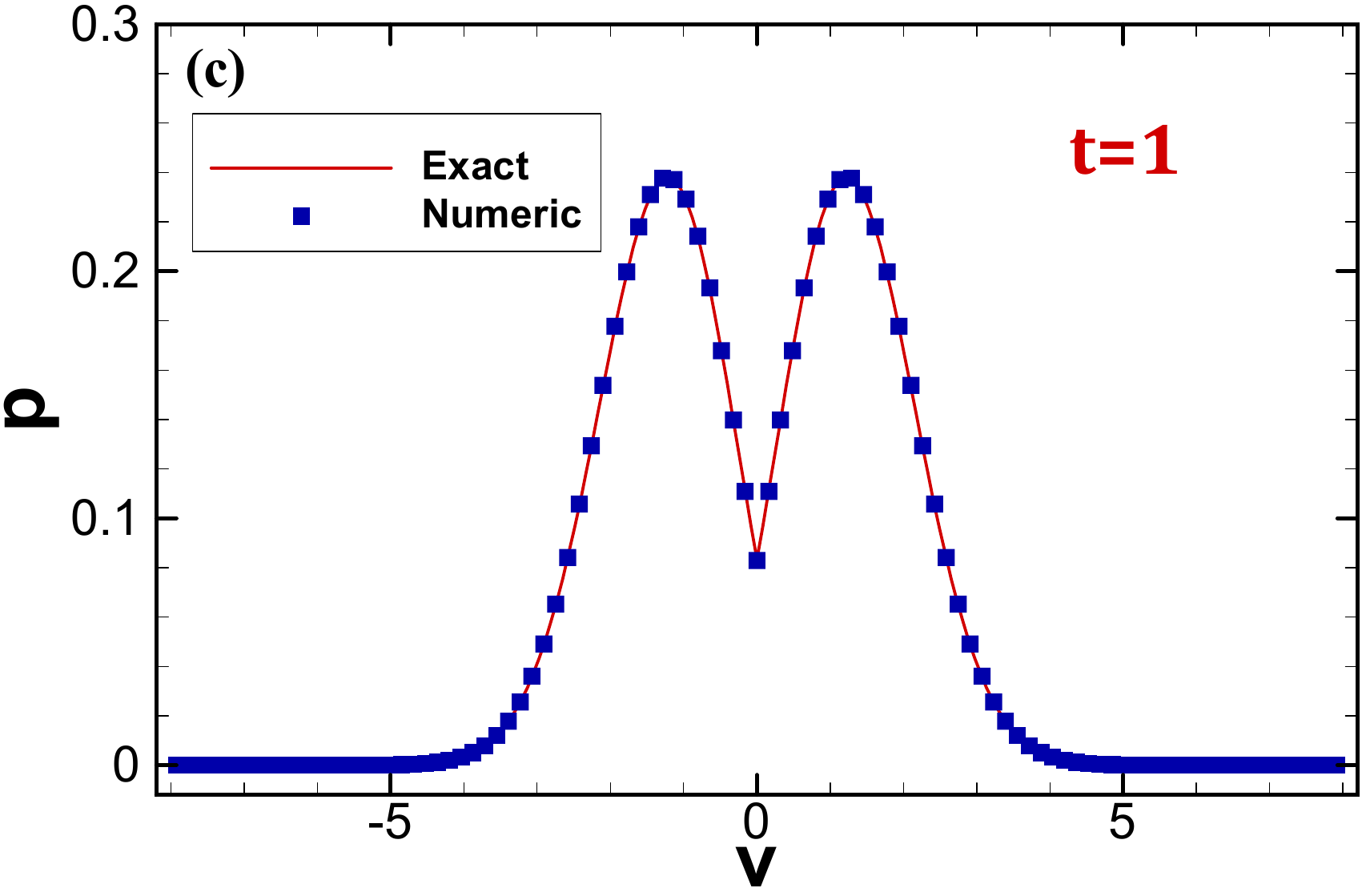}\quad
   \includegraphics[width=0.45\linewidth]{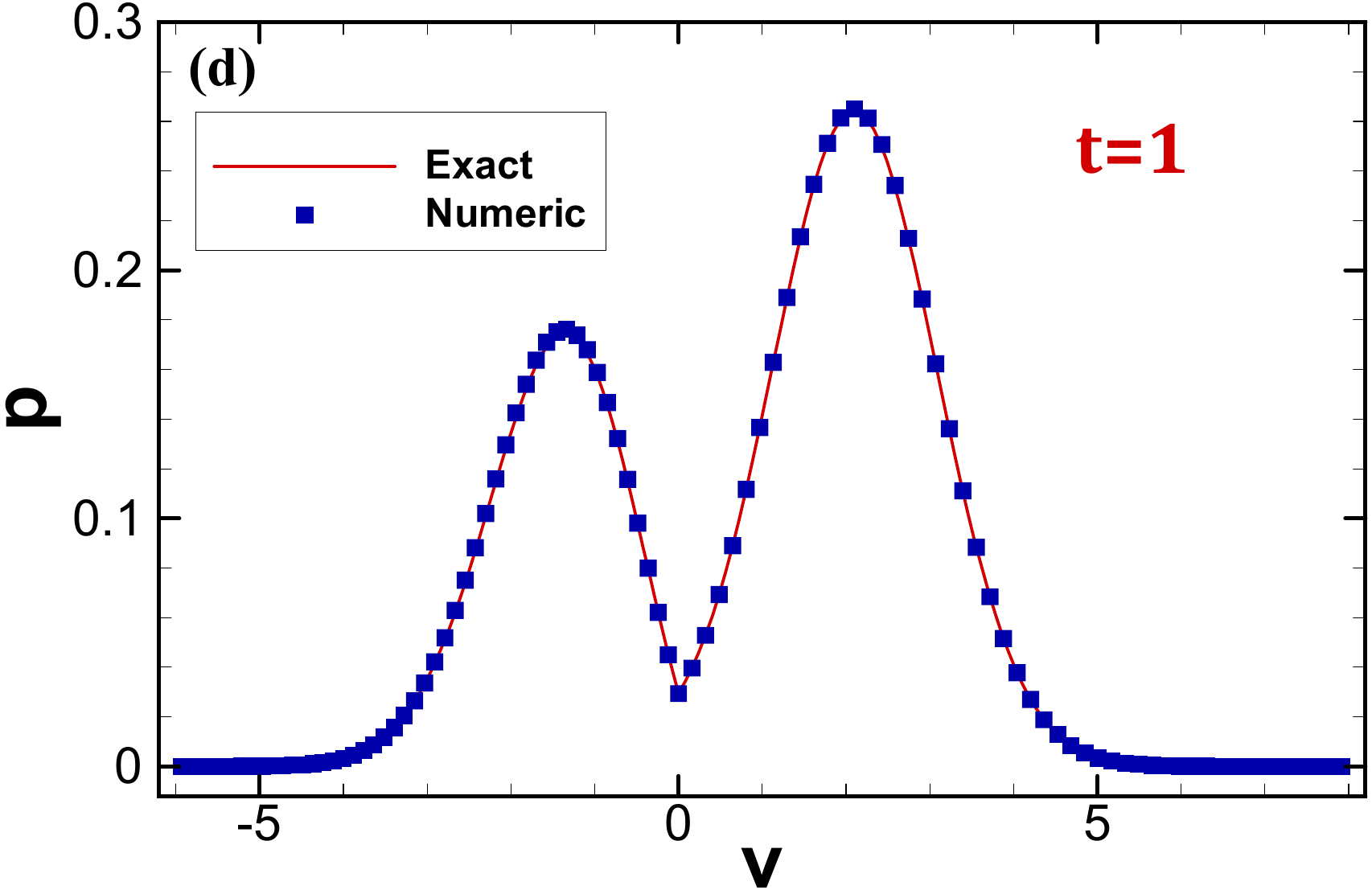}
   \caption{Propagators of the case (\ref{eq_drift_two}) with different values of $\mu_{_L}$ and
   $\mu_{_R}$ at time $t=1$. Here $v_0=0$, $D=0.5$ and $N_1=N_2=50$ are chosen to compute the numerical solutions, which match well with the exact solutions \red{that are obtained by evaluating the convolution integrals appearing in the analytic expression (see Eq.~(5.7) in \cite{Karatzas1984} or Eq.~(42) in \cite{SimpsonKuske2014TwoValued}) numerically by using the routine \emph{NIntegrate} in \emph{Mathematica 8.0}}. (a) $\mu_{_L}=1$ and $\mu_{_R}=2$; (b) $\mu_{_L}=1$ and $\mu_{_R}=-2$;
   (c) $\mu_{_L}=-1$ and $\mu_{_R}=-1$; (d) $\mu_{_L}=-1$ and $\mu_{_R}=-2$.}
   \label{fig_diff_sign}
\end{figure*}

For the case with $\mu_{_L}= \mu_{_R}>0$, this is just the Brownian motion with pure dry friction discussed in Sec.~\ref{sec_4bc}. Here $\mu_{_L}=1$ and $\mu_{_R}=2$ are chosen. The computational domain is set to be $[-3,3]$ and zero boundary conditions are imposed. The result shown in Fig.~\ref{fig_diff_sign}(a) demonstrates the validity of the numerical
method for this case.

\subsection{ Case 2: $ \mu_{_L}>0$ and $\mu_{_R}<0 $ }

In this case, when $\mu_{_L}=-\mu_{_R}=\mu>0$, Eq.~(\ref{ac}) degenerates to
the Brownian motion with a constant drift (see Sec.~\ref{sec_4aa}).
For other cases with $\mu_{_L}\neq -\mu_{_R}$, it is expected that
the propagator is nonsmooth at $v=0$. Here we consider the case with $\mu_{_L}=1$ and $\mu_{_R}=-2$.
The computational domain is chosen to be $[-4,8]$, a zero current condition is set at the left boundary and no boundary condition is specified at the right. The numerical result as shown in Fig.~\ref{fig_diff_sign}(b) agrees with the exact solution, as expected.

\subsection{ Case 3: $\mu_{_L}<0$ and $\mu_{_R}<0$ }

In this case, no boundary condition is needed according to the signs of $\mu_{_L}$ and $\mu_{_R}$. \red{For $v_0=0$} it is expected that the propagator is symmetric for $\mu_{_L}=\mu_{_R}$ and nonsymmetric for $\mu_{_L}\neq \mu_{_R}$. For $\mu_{_L}=\mu_{_R}=-1$, the computational domain $[-8,8]$ is used. The result depicted in Fig.~\ref{fig_diff_sign}(c) shows that the numerical result is consistent with the exact solution. For $\mu_{_L}=-1$ and $\mu_{_R}=-2$, the computational domain $[-6,8]$ is chosen. The result shown in Fig.~\ref{fig_diff_sign}(d) also confirms the validity of the numerical method.
\end{document}